\documentclass[journal]{IEEEtran}
\ifCLASSINFOpdf
\else
\fi

\newtheorem{theorem}{Theorem}[section]
\newtheorem{definition}[theorem]{Definition}
\newtheorem{lemma}[theorem]{Lemma}
\newtheorem{corollary}[theorem]{Corollary}

\newtheorem{exmp}[theorem]{Example}

\newcommand{\EXAMPLE}{\begin{exmp} }
\newcommand{\eoEXAMPLE}{\end{exmp}}

\newcommand{\ignore}[1]{}
\newcommand{\tabincell}[2]{\begin{tabular}{@{}#1@{}}#2 \end{tabular}}
\usepackage{amsmath}
\usepackage{amssymb}
\usepackage{slashbox}
\usepackage[square, comma, sort&compress, numbers]{natbib}

\begin{document}
%
\title{Frequency hopping sequences with optimal partial Hamming correlation}
%
%
%

\author{Jingjun Bao and ~Lijun Ji
\thanks{This work was supported by the NSFC under Grants 11222113, 11431003, and a project funded by the priority academic program development of Jiangsu higher education institutions.}
\thanks{ J. Bao and L. Ji are with the Department of Mathematics, Soochow University, Suzhou 215006, P. R. China. E-mail:  baojingjun@hotmail.com; jilijun@suda.edu.cn.}
}

%
%

\markboth{}%
{Shell \MakeLowercase{\textit{et al.}}: Bare Demo of IEEEtran.cls for Journals}
%



\maketitle

\begin{abstract}
Frequency hopping sequences (FHSs) with favorable partial Hamming correlation properties have important applications in many synchronization and  multiple-access systems. In this paper, we investigate constructions of FHSs and FHS sets with optimal partial Hamming correlation. We first establish a correspondence between  FHS sets with optimal partial Hamming correlation and multiple partition-type balanced nested cyclic difference packings with a special property. By virtue of this correspondence, some FHSs and FHS sets with optimal partial Hamming correlation are constructed from various combinatorial structures such as cyclic difference packings, and cyclic relative difference families. We also describe a direct construction and two recursive constructions for FHS sets with optimal partial Hamming correlation. As a consequence, our constructions yield new FHSs and FHS sets with optimal partial Hamming correlation.
\end{abstract}

\begin{IEEEkeywords}
Frequency hopping sequences (FHSs), partial Hamming correlation, partition-type cyclic difference packings, cyclic relative difference families, cyclotomy.
\end{IEEEkeywords}

%
\IEEEpeerreviewmaketitle

\section{Introduction}
%
%
%
%

\IEEEPARstart{F}{requency} hopping (FH) multiple-access is widely used in the modern communication systems such as ultrawideband (UWB), military communications, Bluetooth and so on, for example, \cite{B2003}, \cite{FD1996},
\cite{YG2004}. In FH  multiple-access communication systems, frequency hopping sequences are employed to specify the frequency on which each sender transmits a message at any given time. An important component of FH spread-spectrum systems is a family of sequences having good correlation properties for sequence length over suitable number of available frequencies. The optimality of correlation property is usually measured according to the well-known Lempel-Greenberger bound and Peng-Fan bounds. During these decades, many algebraic or combinatorial constructions for FHSs or FHS sets meeting these bounds have been proposed,
see \cite{CJ2005}-\cite{CY2010}, \cite{DMY2007}-\cite{DY2008}, \cite{FMM2004}, \cite{GFM2006}-\cite{GMY2009}, \cite{YTUP2011}-\cite{ZTPP2011}, and the references therein.

Compared with the traditional periodic Hamming correlation, the partial Hamming correlation of FHSs is much less well studied. Nevertheless, FHSs with good partial Hamming correlation properties are important for certain application scenarios where an appropriate window length shorter than the total period of the sequences is chosen to minimize the synchronization time or to reduce the hardware complexity of the FH-CDMA receiver \cite{EJHS2004}. Therefore, for these situations, it is necessary to consider the partial Hamming correlation rather than the full-period Hamming correlation.

In 2004, Eun et al. \cite{EJHS2004} generalized the Lempel-Greenberger bound on the periodic Hamming autocorrelation to the case of partial Hamming autocorrelation, and obtained a class of FHSs with optimal partial autocorrelation \cite{US1998}. In 2012, Zhou et al. \cite{ZTNP2012} extended the Peng-Fan bounds on the periodic Hamming correlation of FHS sets to the case of partial Hamming correlation. Based on $m$-sequences, Zhou et al. \cite{ZTNP2012} constructed both individual FHSs and FHS sets with optimal partial Hamming correlation. Very recently, Cai et al. \cite{CZYT2014} improved lower bounds on partial Hamming correlation of FHSs and FHS sets, and based on generalized cyclotomy, they constructed FHS sets with optimal partial Hamming correlation.

\newcounter{mytempeqncnt}
\begin{figure*}[!t]
\normalsize
\setcounter{mytempeqncnt}{\value{equation}}

\centerline{\footnotesize KNOWN AND NEW FHSs WITH OPTIMAL PARTIAL HAMMING CORRELATION }
\vspace{0.1cm}
\begin{center}
\begin{tabular}{|c|c|c|c|c|c|}
\hline
Length & \tabincell{c} {Alphabet \\ size}& \tabincell{c}{$H_{max}$ over correlation\\ windows of length L}& \tabincell{c} {Number of \\ sequences} & Constraints & Source \\
\hline
\tabincell{c}{$q^2-1$\\ } & $q$ & $\left\lceil \frac{L}{q+1}\right\rceil$ & 1 & \tabincell{c}{\\  \\} & \cite{EJHS2004} \\ \hline
\tabincell{c}{$q^m-1$\\ } & $q^{m-1}$ & $\left\lceil \frac{L(q-1)}{q^m-1}\right\rceil$ & 1 & \tabincell{c}{\\  \\} & \cite{ZTNP2012} \\ \hline
\tabincell{c}{$\frac{q^m-1}{d}$} & $q^{m-1}$ & $\left\lceil \frac{L(q-1)}{q^m-1}\right\rceil$ & $d$ & \tabincell{c}{ $d|(q-1)$,\\ gcd$(d,m)=1$ }& \cite{ZTNP2012} \\ \hline
\tabincell{c}{$ev$} & $v$ & $\left\lceil \frac{L}{v}\right\rceil$ & $f$ & \tabincell{c}{\\  \\} & \cite{CZYT2014} \\ \hline
\tabincell{c}{$2v$\\ } & $v$ & $\left\lceil \frac{L}{v}\right\rceil$ & 1 & \tabincell{c}{\\  \\} & Theorem \ref{ strictly optimal 2u} \\ \hline
\tabincell{c}{$2v+1$\\ }  & $v$ & $\left\lceil \frac{L}{v}\right\rceil$ & 1 & \tabincell{c}{\\  \\} & Theorem  \ref{n} \\ \hline
\tabincell{c}{$2v$ } & $\frac{2v+1}{3}$ & $\left\lceil \frac{L}{v} \right\rceil$ & 1 & $p_i$$\equiv 1 \pmod {12}$ is a prime & Theorem \ref{ A FHS-CRDF} \\ \hline
\tabincell{c}{$8v$ } & $\frac{8v+1}{3}$ & $\left\lceil \frac{L}{ 4v } \right\rceil$ & 1 & $p_i$ $\equiv 1 \pmod {6}$ is a prime & Theorem \ref{ A FHS-CRDF} \\ \hline
\tabincell{c}{$32v$} & $\frac{32v+1}{3}$ & $\left\lceil \frac{L}{ 16v } \right\rceil$ & 1  &$p_i$ $\equiv 1 \pmod {12}$ is a prime & Theorem \ref{ A FHS-CRDF} \\ \hline
\tabincell{c}{$3v$ } & $\frac{3v+1}{4}$ & $\left\lceil \frac{L}{ v } \right\rceil$ & 1 & $p_i$ $\equiv 1 \pmod {4}$ is a prime & Theorem \ref{ A FHS-CRDF} \\ \hline
\tabincell{c}{$4v$ } & $\frac{4v+2}{3}$ & $\left\lceil \frac{L}{ 2v } \right\rceil$ & 1 & $p_i$ $\equiv 7 \pmod {12}$ is a prime& Theorem \ref{ A FHS} \\ \hline
\tabincell{c}{$6v$ } & $2v+1$ & $\left\lceil \frac{L}{3v } \right\rceil$ & 1 & $p_i$ $\equiv 5 \pmod {8}$ is a prime & Theorem \ref{ A FHS} \\ \hline
\tabincell{c}{$ p(p^m-1)$ } & $p^{m}$ & $\left\lceil\frac{L}{p^{m}-1}\right\rceil$ & $p^{m-1}$ & \tabincell{c}{ \\ $m\geq 2$ \\  } & Theorem \ref{optimal p} \\ \hline
\tabincell{c}{\\ $evw$ \\ \\ } & $(v-1)w+\frac{ew}{r}$& $\left\lceil \frac{L}{vw}\right\rceil$ & $f$ &  \tabincell{c}{ $q_1\geq p_1>2e$, \\$v>e(e-2)$ and $gcd(w,e)=1$ }  & Corollary \ref{euv} \\ \hline
\tabincell{c}{$v\frac{q^m-1}{d}$ } & $vq^{m-1} $ & $\left\lceil \frac{(q-1)L}{v(q^{m}-1)}\right\rceil$ & $d$ &\tabincell{c}{ $m>1$, and $q^m \leq p_1$\\$\frac{q^m-1}{d}|p_i-1$ for $1\leq i \leq s$  }&  Corollary \ref{d,m} \\ \hline
\end{tabular}
\end{center}

\hspace{1cm} $q$ is a prime power;

\hspace{1cm} $v$ is an integer with prime factor decomposition $v=p_1^{m_1}p_2^{m_2}\cdots p_s^{m_s}$ with $p_1<p_2<\ldots <p_s$;

\hspace{1cm} $e,f$ are integers such that $e>1$ and $e|gcd(p_1-1,p_2-1,\ldots,p_s-1)$, and $f=\frac{p_1-1}{e}$;

\hspace{1cm} $w$ is an integer with prime factor decomposition $w=q_1^{n_1}q_2^{n_2}\cdots q_t^{n_t}$ with $q_1<q_2<\ldots <q_t$;

\hspace{1cm} $r$ is an integer such that $r>1$ and $r|gcd(e,q_1-1,q_2-1,\ldots,q_t-1)$;

\hspace{1cm} $p$ is a prime;

\hspace{1cm} $d,m$ are positive integers.
\setcounter{equation}{\value{mytempeqncnt}}
\vspace*{4pt}
\end{figure*}

In this paper, we present some constructions for FHSs and FHS sets with optimal partial Hamming correlation. First of all, we give combinatorial characterizations of FHSs and FHS sets with optimal partial Hamming correlation. Secondly, by employing partition-type balanced nested cyclic difference packings, cyclic relative difference families, and cyclic relative difference packings, we obtain some FHSs and FHS sets with optimal partial Hamming correlation. Finally, we present two recursive constructions for FHS sets, which increase their lengths and alphabet sizes, and preserve their optimal partial Hamming correlations. Our constructions yield optimal FHSs and FHS sets with new and flexible parameters not covered in the literature. The parameters of FHSs and FHS sets with optimal partial Hamming correlation from the known results and the new ones are listed in the table.

The remainder of this paper is organized as follows. Section II introduces the known bounds on the partial Hamming correlation of FHSs and FHS sets. Section III presents combinatorial characterizations of FHSs and FHS sets with optimal partial Hamming correlation. Section IV gives some combinatorial constructions of FHSs and FHS sets with optimal partial Hamming correlation by using partition-type balanced nested cyclic difference packings, cyclic relative difference families and  cyclotomic classes. Section V presents a direct construction of FHS sets with optimal partial Hamming correlation. Section VI presents two recursive constructions of FHS sets with optimal partial Hamming correlation. Section VII concludes this paper with some remarks.
\section{Lower bounds on the partial Hamming correlation of FHSs and FHS sets} %
\label{pre}                     %

In this section, we introduce some known lower bounds on the partial Hamming correlation of FHSs and FHS sets.

For any positive integer $l\geq2$, let $F=\{f_0, f_1,\ldots, f_{l-1}\}$ be a set of $l$ available frequencies, also called an {\em alphabet}. A sequence $X=\{x(t)\}_{t=0}^{n-1}$ is called a {\em frequency hopping sequence} (FHS) of
length $n$ over $F$ if $x(t)\in F$ for all $0\leq t\leq n-1$. For any two FHSs  $X=\{x(t)\}_{t=0}^{n-1}$ and  $Y=\{y(t)\}_{t=0}^{n-1}$ of length $n$ over $F$, the {\em partial Hamming correlation} function of $X$ and $Y$ for a correlation window length $L$ starting at $j$ is defined by
\begin{equation}
\label{Correlation}
H_{X,Y}(\tau;j|L)=\sum_{t=j}^{j+L-1}h[x(t), y(t+\tau)], 0\leq \tau < n,
\end{equation}
where $L,j$ are integers with $1\leq L\leq n$, $0\leq j< n$, $ h[a,b]=1$ if $a=b$ and $0$ otherwise, and the addition is performed modulo $n$. In particular, if $L=n$, the partial Hamming correlation function defined in ({\ref{Correlation}}) becomes the {\em conventional periodic  Hamming correlation} \cite{LG1974}.
If $x(t)=y(t)$ for all $0\leq t \leq n-1$, i.e., $X=Y$, we call $H_{X,X}(\tau;j|L)$ the {\em partial Hamming autocorrelation} of $X$; otherwise, we say $H_{X,Y}(\tau;j|L)$ the {\em partial Hamming cross-correlation} of $X$ and $Y$. For any two distinct sequences $X,Y$ over $F$ and given integer $1\leq L\leq n$, we define
$$H(X;L)=\max\limits_{0\leq j < n}\max\limits_{1\leq \tau < n}\{H_{X,X}(\tau;j|L)\}$$
and
$$H(X,Y;L)=\max\limits_{0\leq j < n}\max\limits_{0\leq \tau < n}\{H_{X,Y}(\tau;j|L)\}.$$
That is, $H(X;L)$ denotes the maximum partial Hamming autocorrelation of $X$ along with an arbitrary correlation window of length $L$, and $H(X,Y;L)$  denotes the maximum partial Hamming cross-correlation of $X$ and $Y$ along with an arbitrary correlation window of length $L$.

For any FHS of length $n$ over an alphabet of size $l$ and each window length $L$ with $1\leq L\leq n$, Eun et al. \cite{EJHS2004} derived a lower bound:  $H(X;L)\geq \left\lceil \frac{L}{n}\cdot\frac{(n-\epsilon)(n+\epsilon-l)}{l(n-1)}\right\rceil$, where $\epsilon$ is the least nonnegative residue of $n$ modulo $l$, which is a generalization of the Lempel-Greenberger bound \cite{LG1974}. Recently, such a lower bound was improved by Cai et al. \cite{CZYT2014} as follows.

\begin{lemma}(\cite{CZYT2014})\label{FHS BOUND}
Let $X$ be an FHS of length $n$ over an alphabet of size $l$. Then, for each window length $L$ with $1\leq L\leq n,$
\begin{equation}
\label{Bound 4}
H(X;L)\geq \left\lceil\frac{L}{n}\left\lceil \frac{(n-\epsilon)(n+\epsilon-l)}{l(n-1)} \right\rceil \right\rceil,
\end{equation}
where $\epsilon$ is the least nonnegative residue of $n$ modulo $l$.
\end{lemma}

Recall that the correlation window length may change from case to case according to the channel conditions in practical systems. Hence, it is very desirable that the involved FHSs have optimal partial Hamming correlation for any window length. The following definition is originated from the terminology of strictly optimal FHSs in \cite{EJHS2004}.

\begin{definition} \rm
\label{d21-2}
Let $X$ be an FHS of length $n$ over an alphabet $ F$. It is said to be {\em strictly optimal} or an {\em FHS with optimal partial Hamming correlation} if the bound in Lemma \ref{FHS BOUND} is met for an arbitrary correlation window length $L$ with $1 \leq L \leq n$.
\end{definition}

When $L=n$, the bound in Lemma \ref{FHS BOUND} is exactly the Lempel-Greenberger bound \cite{LG1974}. It is clear that each strictly optimal FHS is also optimal with respect to the Lempel-Greenberger bound, but not vice versa \cite{EJHS2004}.

Let $S$ be a set of $M$ FHSs of length $n$ over an alphabet $ F$ of size $l$. For any given correlation window length $L$, the maximum nontrivial partial Hamming correlation $H(S;L)$ of the sequence set $S$ is defined by
$$H(S;L)=\max \{\max\limits_{X\in{S}}H(X;L),\max\limits_{X,Y\in S,\ X\neq Y}H(X,Y;L)\}.$$

Throughout this paper, we use $(n,M,\lambda;l)$ to denote a set $S$ of $M$ FHSs of length $n$ over an alphabet $F$ of size $l$, where $\lambda=H(S;n)$, and we use $(n,\lambda;l)$ to denote an FHS $X$ of length $n$ over an alphabet $F$ of size $l$, where $\lambda=H(X;n)$.

When $L=n$, Peng and Fan \cite{PF2004} described the following bounds on $H(S;n)$, which take into consideration the number of sequences in the set $S$.

\begin{lemma}(\cite{PF2004})\label{set}
Let $S$ be a set of $M$ sequences of length $n$ over an alphabet $F$ of size $l$.
Define $I=\left\lfloor \frac{nM}{l}\right\rfloor$. Then

$$H(S;n)\geq \left\lceil \frac{(nM-l)n}{(nM-1)l}\right\rceil$$
and
$$H(S;n)\geq \left\lceil \frac{2InM-(I+1)Il}{(nM-1)M}\right\rceil.$$
\end{lemma}

In 2012, Zhou et al. \cite{ZTNP2012} extended the Peng-Fan bounds to the case of the partial Hamming correlation. They obtained $ H(S;L)\geq \left\lceil \frac{L}{n}\cdot \frac{(nM-l)n}{(nM-1)l}\right \rceil $ and $ H(S;L)\geq \left\lceil \frac{L}{n} \cdot \frac{2InM-(I+l)Il}{(nM-1)M} \right\rceil.$
Recently, such lower bounds were improved by Cai et al. \cite{CZYT2014}.

\begin{lemma}(\cite{CZYT2014}) \label{optimal set}
Let $S$ be a set of $M$ FHSs of length $n$ over $F$ of size $l$.
Define $I=\lfloor \frac{nM}{l}\rfloor$. Then, for an arbitrary window length $L$ with $1\leq L\leq n$,
\begin{equation}
\label{Bound 5}
H(S;L)\geq \left\lceil \frac{L}{n} \left\lceil\frac{(nM-l)n}{(nM-1)l}\right\rceil\right \rceil
\end{equation}
and
\begin{equation}
\label{Bound 6}
H(S;L)\geq \left\lceil \frac{L}{n} \left\lceil\frac{2InM-(I+1)Il}{(nM-1)M}\right\rceil \right\rceil.
\end{equation}
\end{lemma}

Parallel to the notion of strictly optimal individual FHSs, Cai et al. gave the following definition of strictly optimal FHS sets in \cite{CZYT2014}.

\begin{definition} \rm
\label{d111}
An FHS set $S$ is said to be {\em strictly optimal} or an {\em FHS set with optimal partial Hamming correlation} if one of the bounds in Lemma \ref{optimal set} is met for an arbitrary correlation window length $L$ with $1 \leq L \leq n.$
\end{definition}

Note that, when $L=n$, the bounds in Lemma \ref{optimal set} are exactly the Peng-Fan bounds. It turns out that each strictly optimal FHS set is also optimal with respect to the Peng-Fan bounds, but not vice versa.

\section{ Combinatorial characterizations }     %
\subsection{A combinatorial characterization of Strictly optimal FHSs} %

A convenient way of viewing FHSs is from a set-theoretic perspective. In 2004, Fuji-Hara et al. \cite{FMM2004} revealed a connection between FHSs and partition-type cyclic difference packings. In this subsection, we give a combinatorial characterization of strictly optimal FHSs.

Throughout this paper we always assume that $\mathbf{I}_l=\{0,1,2,\ldots,l-1\}$ and $\mathbb{Z}_n$ is the residual-class ring of integers modulo $n$. For integers $s,n$, the notation $s|n$ means that $n$ is divisible by $s$. For $A\subset \mathbb{Z}_n$, $\tau\in \mathbb{Z}_n$ and a positive integer, $A+\tau$ is defined to be $\{a+\tau : a\in A\}$ and $\lambda\cdot[A]=\bigcup\limits_{i=0}^{\lambda-1}A$, where $\bigcup$ is the multiset union.

An $(n,\lambda;l)$-FHS, $X=(x(0),x(1),\ldots,x(n-1))$, over a frequency library $F$ can be interpreted as a set ${\cal B}$ of $l$ sets $B_0,B_1,\ldots,B_{l-1}$ such that each set $B_i$ corresponds to frequency $i\in {F}$ and the elements in each set $B_i$ specify the position indices in the FHS $X$ at which frequency $i$ appears. The correlation property can be rephrased in this set-theoretic framework.
As stated in \cite[Lemma 2.1]{FMM2004}, the set ${\cal B}$ is a partition of $\mathbb{Z}_n$ and $H_{X,X}(\tau;0|n)=\sum \limits_{i=0}^{l-1}|B_i\cap (B_i+\tau)|\leq \lambda$ for $1\leq \tau<n$, i.e.,  ${\cal B}$ satisfies the following property in terms of differences: each $\tau\in \mathbb{Z}_n\setminus \{0\}$ can be represented as a difference $b'-b$, $b,b'\in B_i$, $0\leq i\leq l-1$, in at most $\lambda$ ways. This observation reveals a connection between FHSs and combinatorial structures called   partition-type cyclic difference packings.

Associated with a non-empty subset $B \subseteq \mathbb{Z}_n$, the difference list of $B$ from combinatorial design theory is defined to be the multiset
  $$ \Delta (B) = \left\{a-b:~ a, b \in B~ \mbox {and}~  a \ne b\right\}.$$
For any family ${\cal B} = \{B_{0}, B_{1}, \ldots, B_{l-1}\}$
 of $l$ non-empty subsets (called {\em base blocks}) of $\mathbb{Z}_n$, define the difference list of ${\cal B}$ to be the union of multisets

 $$ \Delta ({\cal B}) = \displaystyle{\bigcup_{i\in \mathbf{I}_l}\Delta (B_i)}.$$
If the difference list $\Delta({\cal B})$ contains each non-zero residue of $\mathbb{Z}_n$ at most $\lambda$ times, then ${\cal B}$ is said to be an $(n, K, \lambda)$-CDP ({\em cyclic difference packing}), where $K=\{|B_i|: i\in \mathbf{I}_l\}$. When $K=\{k\}$, we simply write $k$ instead of $\{k\}$. The number $l$ of the base blocks in ${\cal B}$ is refereed to as the size of the CDP. If the difference list $\Delta({\cal B})$ contains each non-zero residue of $\mathbb{Z}_n$ exactly $\lambda$ times, then ${\cal B}$ is a {\em  cyclic difference family} and denoted by $(n, K, \lambda)$-CDF.

An $(n,K,\lambda)$-CDP ${\cal B}=\{B_0,B_1,\ldots,B_{l-1}\}$ is called a {\em partition-type cyclic difference packing} if every element of $\mathbb{Z}_n$ is contained in exactly one block in ${\cal B}$.


In 2004, Fuji-Hara et al. \cite{FMM2004} revealed a connection between FHSs and partition-type cyclic difference packings as follows.
\begin{theorem}{\rm(\cite{FMM2004})}\label{FHS=DP}
There exists an $(n,\lambda;l)$-FHS over a frequency library ${F}$ if and only if there exists a partition-type $(n,K,\lambda)$-CDP of size $l$, ${\cal B}=\{B_0,B_1,\ldots,B_{l-1}\}$ over $\mathbb{Z}_n$, where $K=\{|B_i|:~0\leq i\leq l-1\}$.
\end{theorem}

Fuji-Hara et al. \cite{FMM2004} also gave a simplified version of the Lempel-Greenberger bound as follows.

\begin{lemma}(\cite{FMM2004})\label{A optimal}
For an arbitrary $(n,\lambda; l)$-FHS, it holds that
\[
\begin{array}{l}
\lambda \geq
\left\{\begin{array}{ll}
k & {\rm  if} \ \ n\neq l,\ \ {\rm and} \\
0  & {\rm  if} \ \ n= l,\
\end{array}
\right .
\end{array}
\]
where $n=kl+\epsilon,\ 0\leq \epsilon \leq l-1.$ This implies that when $n>l$, the sequence is optimal if $\lambda=k$.
\end{lemma}

\begin{lemma}{\rm(\cite{FMM2004})}\label{OFHS=DF}
Let $n=kl+l-1$ with $k\geq1$. Then there exists an optimal $(n,k;l)$-FHS if and only if there exists a partition-type $(n,\{k,k+1\},k)$-CDF in which $l-1$ blocks are of size $k+1$ and the remaining one is of size $k$.
\end{lemma}

For a $u$-tuple $T=(a_0,a_1,\ldots,a_{u-1})$ over $\mathbb{Z}_n$, the multiset $\Delta_i(T)=\{a_{j+i}-a_j:\ 0\leq j\leq u-1\}$ is called $i$-apart difference list of the tuple, where $1\leq i\leq u$, $j+i$ is reduced modulo $u$, and $a_{j+i}-a_j$ is taken as the least positive residue modulo $n$.

For a partition-type CDP of size $l$ over $\mathbb{Z}_n$, ${\cal B}=\{B_0,B_1,\ldots,B_{l-1}\}$, for $1\leq \tau \leq n-1$, let
\[
\begin{array}{l}
D(\tau)=\{a:~ 0\leq a< n, \{a,\ a+\tau\} \subseteq B, B\in {\cal B}\},\\
\overrightarrow{D}(\tau)=(a_0,a_1,\ldots,a_{u-1}), {\rm where}\ 0\leq a_0<\cdots <a_{u-1}<n\ \\ \hspace{1.3cm} {\rm and}\ \{a_0,a_1,\ldots,a_{u-1}\}=D(\tau),  {\rm and} \\
 d^{\cal B}_i=\min \limits_{1\leq \tau<n}\min \{g: g\in \{n\}\cup \Delta_i(\overrightarrow{D}(\tau))\}, \vspace{0.1cm}\\ \hspace{1.1cm} {\rm for}\ 1\leq i\leq \lambda=\max\{|D(\tau)|: 1\leq \tau <n \},
\end{array}
\]
where $\Delta_i(\overrightarrow{{D}}(\tau))=\emptyset$ if $D(\tau)=\emptyset$ or $i> |D(\tau)|$. We call $\overrightarrow{D}(\tau)$  the orbit cycle of $\tau$ in ${\cal B}$ and $ d^{\cal B}_i$ the minimal $i$-apart distance of all orbits.
Note that $D(-\tau)\equiv D(\tau)+\tau \pmod n$ and $\min \{g: g\in \{n\}\cup\Delta_i(\overrightarrow{D}(\tau))\}=\min \{g: g\in \{n\}\cup \Delta_i(\overrightarrow{D}(-\tau))\}$.

We illustrate the definition of $d^{\cal B}_i$ in the following example.
\begin{exmp}\label{30}
In this example, we construct a partition-type $(30,11,2)$-CDP, ${\cal B}$, with $d^{\cal B}_i=15i$ for $0<i \leq 2$.

 Set
\[
\begin{array}{l}
\vspace{0.2cm}{\cal B}=\{\{1,2,14\},\{3,7,9\},\{4,23,26\}, \{6,13,27\}, \\ \vspace{0.2cm} \ \ \{16,17,29\}, \{18,22,24\},\{8,11,19\}, \{12,21,28\},\\  \ \ \ \ \{0,5\},\{15,20\},\{10,25\}\}.
\end{array}
\]

It is readily checked that ${\cal B}$ is a $(30,11,2)$-CDP. By the fact $\min \{g: g\in \{n\} \cup \Delta_{i}(\overrightarrow{D}(\tau))\}=\min \{g: g\in \{n\} \cup \Delta_{i}(\overrightarrow{D}(-\tau))\}$, we only need to compute $\overrightarrow{D}(\tau)$ for $1\leq \tau \leq 15$, so as to obtain $d_{i}^{\cal B}$.
Simple computation shows that
\[
\begin{array}{l}
\vspace{0.2cm}\overrightarrow{D}(1)=(1,16), \overrightarrow{D}(2)=(7,22), \overrightarrow{D}(3)=(8,23), \\
\vspace{0.2cm}\overrightarrow{D}(4)=(3,18), \overrightarrow{D}(5)=(0,15), \overrightarrow{D}(6)=(3,18), \\
\vspace{0.2cm}\overrightarrow{D}(7)=(6,21), \overrightarrow{D}(8)=(11,26), D(9)=\emptyset, \\
\vspace{0.2cm}D(10)=\emptyset, \overrightarrow{D}(11)=(8,23), \overrightarrow{D}(12)=(2,17), \\
\vspace{0.2cm}\overrightarrow{D}(13)=(1,16), \overrightarrow{D}(14)=(13,28), \overrightarrow{D}(15)=(10,25).\\
\end{array}
\]
Then, $d^{\cal B}_i=15i$ for $0<i \leq 2$. By direct check or by Theorem \ref{optimal character}, the corresponding FHS is a strictly optimal $(30,2;11)$-FHS.
\end{exmp}

We are in a position to give a combinatorial characterization of strictly optimal FHSs.

\begin{theorem}\label{optimal character}
There is a strictly optimal $(n,\left\lfloor\frac{n}{l}\right\rfloor;l)$-FHS with respect to the bound $(\ref{Bound 4})$ if and only if there exists a partition-type $(n,K,\left\lfloor\frac{n}{l}\right\rfloor)$-CDP of size $l$, ${\cal B}=\{B_0,B_1,\ldots,B_{l-1}\}$ over $\mathbb{Z}_n$, such that $d_i^{{\cal B}}\geq \left\lfloor \frac{ni}{\left\lfloor\frac{n}{l}\right\rfloor}\right \rfloor$ for $1\leq i\leq \left\lfloor\frac{n}{l}\right\rfloor$, where $K=\{|B_r|:~0\leq r\leq l-1\}$.
\end{theorem}

\begin{IEEEproof}
Let $X$ be an FHS of length $n$ over $F=\mathbf{I}_l$. Let $B_r=\{t:~ x(t)=r, 0\leq t\leq n-1\}$ for each $r \in F.$ Then $B_r$ is the set of position indices in the sequence $X=(x(0),\ldots,x(n-1))$ at which the frequency $r$ appears. By Theorem \ref{FHS=DP}, the sequence $X$ is an $(n,\left\lfloor\frac{n}{l}\right \rfloor;l)$-FHS over $F$ if and only if ${\cal B}=\{B_0,B_1,\ldots,B_{l-1}\}$ is a partition-type $(n,K,\left\lfloor\frac{n}{l}\right \rfloor)$-CDP of size $l$, where $K=\{|B_r|:~0\leq r\leq l-1\}$. It is left to show that $X$ is strictly optimal with respect to the bound $(\ref{Bound 4})$ if and only if $d_i^{{\cal B}}\geq \left\lfloor \frac{ni}{\left\lfloor\frac{n}{l}\right\rfloor}\right \rfloor$ for $1\leq i\leq \left\lfloor\frac{n}{l}\right\rfloor$.

Denote $\lambda=\left\lfloor\frac{n}{l}\right \rfloor$. By Lemma \ref{FHS BOUND}, $X$ is strictly optimal with respect to bound $(\ref{Bound 4})$ if and only if  $H(X;L)\leq \left\lceil\frac{L\lambda}{n}\right\rceil$ for $1\leq L\leq n$, i.e., for $1\leq i\leq \lambda$, the inequality
$H(X;L)\leq i$ holds for $\left\lfloor\frac{n(i-1)}{\lambda}\right\rfloor <L \leq \left\lfloor\frac{ni}{\lambda}\right\rfloor$. We show that the inequality
$H(X;L)\leq i$ holds for $\left\lfloor\frac{n(i-1)}{\lambda}\right\rfloor <L \leq \left\lfloor\frac{ni}{\lambda}\right\rfloor$ if and only if $d^{{\cal B}}_i\geq \left\lfloor \frac{ni}{\lambda}\right \rfloor$.

By the definition of $H_{X,X}(\tau;j|L)$ and $D(\tau)$, we have $H_{X,X}(\tau;j|L)=|D(\tau;j|L)|$, where $D(\tau;j|L)=D(\tau)\cap \{j,j+1,j+2,\ldots, j+L-1\}$ and the arithmetic $j+t$ $(1\leq t\leq L-1)$ is reduced modulo $n$. Denote $d_i(\tau)=\min \{g: g\in \{n\}\cup\Delta_i(\overrightarrow{D}(\tau))\}$. Clearly, $\max\limits_{0\leq j < n}\{|D(\tau;j|L)|\}\leq i$ if and only if $L\leq d_i(\tau)$. Hence, $H(X;L)=\max\limits_{1\leq \tau < n}\max\limits_{0\leq j < n}\{|D(\tau;j|L)|\}\leq i$ if and only if $L\leq d_i(\tau)$ for $1\leq \tau<n$, i.e., $L\leq d_i^{{\cal B}}$=$\min\{d_i(\tau): 1\leq \tau<n\}$. It follows that $H(X;L)\leq i$ for $\left\lfloor\frac{n(i-1)}{\lambda}\right\rfloor <L \leq \left\lfloor\frac{ni}{\lambda}\right\rfloor$ if and only if $d^{\cal B}_i\geq \left\lfloor \frac{ni}{\lambda}\right \rfloor$. This completes the proof.
\end{IEEEproof}


An $(mg, g, K, \lambda)$-{\em cyclic relative difference packing} (briefly CRDP) is an $(mg,K,\lambda)$-CDP  over $\mathbb{Z}_{mg}$, ${\cal B}=\{B_i:~ i\in I_u \}$, such that no element of $m\mathbb{Z}_{mg}$ occurs in  $\Delta({\cal B})$, where $m\mathbb{Z}_{mg}=\{0,m,\ldots,(g-1)m\}$. The number $u$ of the base blocks in ${\cal B}$ is refereed to as the size of the CRDP. If $\Delta({\cal B})$ contains each element of $\mathbb{Z}_{mg}\setminus m\mathbb{Z}_{mg}$ exactly $\lambda$ times and no element of $m\mathbb{Z}_{mg}$ occurs, then ${\cal B}$ is called an $(mg, g, K, \lambda)$-{\em cyclic relative difference family} (briefly CRDF).


One importance of a CRDP is that we can put an appropriate CDP on its subgroup to derive a new CDP. We state a simple but useful fact in the following lemma.

\begin{lemma}\label{CRDP=>CDP}
Let ${\cal B}$ be an $(mg,g,K,1)$-CRDP of size $u$ over $\mathbb{Z}_{mg}$ relative to $m\mathbb{Z}_{mg}$, whose elements of base blocks, together with $0,m,\ldots,(s-1)m,$ form a complete system of representatives for the cosets of $sm\mathbb{Z}_{mg}$ in $\mathbb{Z}_{mg}$, where $s|g$. Let ${\cal A}$ be a partition-type $(g,K',\frac{g}{s})$-CDP of size $r$ over $\mathbb{Z}_g$  with $d_i^{\cal A}\geq si$ for $1\leq i\leq \frac{g}{s}$. Then there exists a
partition-type $(mg,K\cup K',\frac{g}{s})$-CDP of size $\frac{gu}{s}+r$, ${\cal D}$ such that $d_i^{{\cal D}}\geq ism$ for $1\leq i\leq \frac{g}{s}$.
\end{lemma}

\begin{IEEEproof}
Let ${\cal B}'=\{B+jsm: B\in {\cal B}, 0\leq j<{\frac{g}{s}}\}$, ${\cal A}'=\{mA: A\in {\cal A}\}$ and ${\cal D}={\cal B}'\cup {\cal A}'$.
Since ${\cal B}$ is an $(mg,g,K,1)$-CRDP of size $u$ over $\mathbb{Z}_{mg}$ relative to $m\mathbb{Z}_{mg}$ and all elements of base blocks, together with $0,m,2m,\ldots,(s-1)m,$ form a complete system of representatives for the cosets of $sm\mathbb{Z}_{mg}$ in $\mathbb{Z}_{mg}$, we have that $\Delta({\cal B})\subset \mathbb{Z}_{mg}\setminus m\mathbb{Z}_{mg}$ and ${\cal B}'$ is a partition of $\mathbb{Z}_{mg}\setminus m\mathbb{Z}_{mg}$.
Clearly, $\Delta({\cal B'})$ contains each $\tau\in \Delta({\cal B})$ exactly $\frac{g}{s}$ times and no element of $\mathbb{Z}_{mg}\setminus \Delta({\cal B})$.
Since ${\cal A}$ is a partition-type $(g,K',\frac{g}{s})$-CDP of size $r$ over $\mathbb{Z}_g$, we get $\bigcup\limits_{A\in {\cal A}}\Delta(mA)=m\Delta({\cal A})$ and ${\cal A}'$ is a partition of $m\mathbb{Z}_{mg}$. Hence, ${\cal D}$ is a partition-type $(mg,K\cup K',\frac{g}{s})$-CDP of size $\frac{gu}{s}+r$.

Clearly, if $\tau \in \Delta({\cal B})$, then the orbit cycle $\overrightarrow{D}(\tau)$ of $\tau$ in ${\cal D}$ is of the form $(a_0,a_0+sm,a_0+2sm,\ldots, a_0+mg-sm)$, and if $\tau\in \Delta({\cal A}')$, then the orbit cycle $\overrightarrow{D}(\tau)$ of $\tau$ in ${\cal D}$ is of the form $(mb_0,mb_1,\ldots,mb_{s'-1})$, where $(b_0,b_1,\ldots,b_{s'-1})$ is the orbit cycle of $\frac{\tau}{m}$ in ${\cal A}$, otherwise $D(\tau)=\emptyset$.
Hence, $d^{{\cal D}}_i(\tau)=ism$ for $\tau \in \Delta({\cal B})$, $d^{{\cal D}}_i(\tau)\geq ism$ for $\tau\in \Delta({\cal A}')$ because $d_i^{\cal A}\geq si$ for $1\leq i\leq \frac{g}{s}$, and $d^{{\cal D}}_i(\tau)=mg$ for $\tau \notin \Delta({\cal D})$, where $d^{{\cal D}}_i(\tau)=\min\{g:~g\in \{mg\}\cup \Delta_i(\overrightarrow{D}(\tau))\}$. It follows that $d_i^{{\cal D}}=\min\{d^{{\cal D}}_i(\tau):0<\tau <mg\}=ism$ for $1\leq i\leq \frac{g}{s}$. Therefore, ${\cal D}$ is the required partition-type $(mg,K\cup K',\frac{g}{s})$-CDP. This completes the proof.
\end{IEEEproof}

Let ${\cal B}$ be an $(mg,g,k,1)$-CRDF with $k-1|g$, if all elements of base blocks, together with $0,m,\ldots,(s-1)m,$ form a complete system of representatives for the cosets of $sm\mathbb{Z}_{mg}$ in $\mathbb{Z}_{mg}$, where $s=\frac{g}{k-1}$, then this $(mg,g,k,1)$-CRDF is said to be {\em resolvable}. There are several constructions for resolvable $(mg,g,k,1)$-CRDFs in \cite{BZ2001}-\cite{BZ1998}.

For example, the set ${\cal B}'=\{\{1,2,14\}$, $\{3,7,9\}$, $\{4,23,26\}$, $\{6,13,27\}\}$ is a resolvable $(30,6,3,1)$-CRDF over $\mathbb{Z}_{30}$. It is easy to see that ${\cal A}=\{\{0,1\},\{3,4\},\{2,5\}\}$ is a $(6,2,2)$-CDP with $d_i^{\cal A}\geq 3i$ for $1\leq i\leq 2$. By Lemma \ref{CRDP=>CDP}, we also obtain a $(30,\{2,3\},2)$-CDP of size 11 such that $d_i=15i$ for $1\leq i\leq 2$, which is the same as that in Example \ref{30}.

By virtue of the combinatorial characterization of strictly optimal FHSs, the existence of a strictly optimal $(kv,k;\frac{kv+1}{k+1})$-FHS is equivalent to the existence of a resolvable $(kv,k,k+1,1)$-CRDF.

\begin{theorem}\label{PHFS-CRDF}
Let $k$ and $v$ be positive integers with $k+1|v-1$. Then there exists a strictly optimal $(kv,k;\frac{kv+1}{k+1})$-FHS if and only if there exists a resolvable  $(kv,k,k+1,1)$-CRDF.
\end{theorem}
\begin{IEEEproof}: By Lemma \ref{OFHS=DF} and Theorem \ref{optimal character}, there is a strictly optimal $(kv,k;\frac{kv+1}{k+1})$-FHS if and only if there is a partition-type $(kv,\{k,k+1\},k)$-CDF over $\mathbb{Z}_{kv}$ with $ d_i\geq vi$ for $1\leq i\leq k$, in which $\frac{kv-k}{k+1}$ blocks are of size $k+1$ and the remaining one is of size $k$. We shall prove that such a partition-type $(kv,\{k,k+1\},k)$-CDF exists if and only if there is a resolvable $(kv,k,k+1,1)$-CRDF relative to $v\mathbb{Z}_{kv}=\{0,v,\ldots, (k-1)v\}$.

Let ${\cal B}$ be such a partition-type $(kv,\{k,k+1\},k)$-CDF where ${\cal B}=\{B_0,B_1,\ldots,B_{(kv-k)/(k+1)}\}$. By the definition of CDF, for $1\leq \tau\leq kv-1$, each difference $\tau$ occurs exactly $k$ times, thus $\overrightarrow{D}(\tau)=(a_0,a_1,\ldots,a_{k-1})$ is a $k$-tuple vector of distinct elements. Clearly, $d_1(\tau)=\min\{g: g\in \{kv\} \cup \Delta_1(\overrightarrow{D}(\tau))\}\leq \lfloor\frac{kv}{k}\rfloor= v$. Since $d_i \geq iv$, we have $d_1(\tau)= v$. This leads to $a_{j+1}=a_j+v$ for $0\leq j\leq k-1$. Hence, for any pair $\{a,b\}$ contained in some base block of ${\cal B}$, the pair $\{a+v,b+v\}$ is also contained in a base block of ${\cal B}$. Since ${\cal B}$ is a partition-type CDF, the set $B_j+v$ is also a base block of ${\cal B}$ and one base block of ${\cal B}$ is of the form $v\mathbb{Z}_{kv}+t$. Without loss of generality, let $B_{(kv-k)/(k+1)}=v\mathbb{Z}_{kv}+t$ and $B_j+v=B_{j+\frac{v-1}{k+1}}$ for $0\leq j< (kv-k)/(k+1)$, where $j+\frac{v-1}{k+1}$ is reduced modulo $\frac{kv-k}{k+1}$.
It is easy to see that $\{B_0-t,B_1-t,\ldots,B_{(v-k-2)/(k+1)}-t\}$ is the set of base blocks of a resolvable $(kv,k,k+1,1)$-CRDF relative to $v\mathbb{Z}_{kv}$.

Conversely, suppose that there is a resolvable $(kv,k,k+1,1)$-CRDF over $\mathbb{Z}_{kv}$ relative to $v\mathbb{Z}_{kv}$. Then $s=\frac{k}{k}=1$. Since ${\cal A}=\{\mathbb{Z}_k\}$ is the trivial partition-type $(k,k,k)$-CDF with $d_i^{{\cal A}}=i$ for $1\leq i\leq k$,  applying Lemma \ref{CRDP=>CDP} yields a partition-type $(kv,\{k,k+1\},k)$-CDF of size $\frac{kv+1}{k+1}$, ${\cal D}$, such that $d_i^{{\cal D}}\geq vi$ for $1\leq i\leq k$.  This completes the proof.
\end{IEEEproof}

\subsection{A combinatorial characterization of strictly optimal FHS sets} %

In 2009, Ge et al. \cite{GMY2009} revealed a connection between FHS sets and families of partition-type balanced nested cyclic difference packings. In this subsection, we extend the combinatorial approach used in Section III-A for designing strictly optimal FHSs to give a combinatorial characterization of strictly optimal FHS sets.

Let $S$ be an $(n,M,\lambda;l)$-FHS set over $F=\mathbf{I}_l$. For $X\in S$, let $B_r^X=\{t:~ x(t)=r, 0\leq j\leq n-1\}$ for each $r \in F$. Then $B_r^X$ is the set of position indices in the sequence $X=(x(0),\ldots,x(n-1))$ at which the frequency $r$ appears.
For two distinct FHSs $X,Y\in S$, it is easy to see that the Hamming cross-correlation $H_{X,Y}(\tau;0|n)=\sum \limits_{i=0}^{l-1}|(B_i^X+\tau)\cap B_i^Y|\leq \lambda$ for $0\leq \tau<n$, i.e.,  $\{B_i^X: i\in \mathbf{I}_l\}$ and $\{B_i^Y: i\in \mathbf{I}_l\}$ satisfy the following property in terms of external differences: each $\tau\in \mathbb{Z}_n$ can be represented as a difference $b'-b$, $(b,b')\in B_i^X\times B_i^Y$, $0\leq i\leq l-1$, in at most $\lambda$ ways. This observation reveals a connection between FHS sets and combinatorial structures called families of partition-type balanced nested cyclic difference packings.

Let $A,B$ be two subsets of $\mathbb{Z}_v$.  {\em The list of external difference} of ordered pair $(A,B)$ is the multiset $$\Delta_E(A,B)=\{y-x:~(x,y)\in A\times B\}.$$  Note that the list of external difference $\Delta_E(A,B)$ may contain zero. For any residue $\tau\in \mathbb{Z}_v$, the number of occurrences of $\tau$ in $\Delta_E(A,B)$ is clearly equal to $|(A+\tau)\cap B|$.

Let ${\cal B}_j, 0\leq j\leq M-1$, be a collection of $l$ subsets $B_0^j, \ldots,B_{l-1}^j$ of $\mathbb{Z}_n$, respectively. The list of external difference of ordered pair $({\cal B}_j,{\cal B}_{j'})$, $0\leq j\neq j'<M$, is the union of multisets
$$\Delta_E({\cal B}_j,{\cal B}_{j'})= \bigcup\limits_{i\in \mathbf{I}_l} \Delta_E(B_i^j,B_i^{j'}).$$
If each ${\cal B}_j$ is an $(n,K_j,\lambda)$-CDP of size $l$, and $\Delta_E({\cal B}_j, {\cal B}_{j'})$
contains each residue of $\mathbb{Z}_n$ at most $\lambda$ times for $0\leq j\neq j'<M$, then the set $\{{\cal B}_0, \ldots, {\cal B}_{M-1}\}$ of CDPs is said to be {\em balanced nested} with index $\lambda$ and denoted by $(n,\{K_0,\ldots, K_{M-1}\},\lambda)$-BNCDP. If each ${\cal B}_j$ is a partition-type CDP for $0\leq j<M$, then the $(n,\{K_0,\ldots, K_{M-1}\},\lambda)$-BNCDP is called {\em partition-type}. For convenient, the number $l$ of the base blocks in ${\cal B}_j$ is also said to be the size of the BNCDP.

In 2009, Ge et al. \cite{GMY2009} revealed a connection between FHS sets and partition-type BNCDPs as follows.

\begin{theorem}{\rm(\cite{GMY2009})}\label{FHS set=DPs}
There exists an $(n,M,\lambda;l)$-FHS set over a frequency library $F$ if and only if there exists a partition-type  $(n,\{K_0,K_1,\ldots,K_{M-1}\},\lambda)$-BNCDP of size $l$.
\end{theorem}

 Let ${\cal B}=\{{\cal B}_X: X\in S\}$ be a family of $M$ partition-type CDPs of size $l$ over $\mathbb{Z}_n$, where ${\cal B}_X=\{B_0^X,B_1^X,\ldots,B_{l-1}^X\}, X \in S $. For two distinct partition-type CDPs ${\cal B}_X$, ${\cal B}_Y$, and for $0\leq \tau \leq n-1$, let
\[
\begin{array}{l}
\vspace{0.1cm}D_{(X,Y)}(\tau)=\{a:~ 0\leq a <n, \ (a,\ a+\tau) \in B_i^X\times B_i^Y, \\ \hspace{1.3cm}\vspace{0.1cm}{\rm for\ some}\ i\in \mathbf{I}_l\},\\
\vspace{0.1cm}\overrightarrow{D_{(X,Y)}}(\tau)=(a_0,a_1,\ldots,a_{u-1}),\\ \hspace{1.5cm} {\rm where}\ 0\leq a_0<\cdots <a_{u-1}<n\ \\ \hspace{1.5cm} {\rm and}\ \{a_0,a_1,\ldots,a_{u-1}\}=D_{(X,Y)}(\tau), \\
 \end{array}
\]
\[
\begin{array}{l}
\vspace{0.1cm} d_i^{(X,Y)}=\min\limits_{0\leq \tau<n}\min \{g: g\in \{n\}\cup\Delta_i(\overrightarrow{D_{(X,Y)}}(\tau))\}, \\
\vspace{0.1cm}\hspace{1.3cm} {\rm for}\ 0 \leq i\leq \max\{|D_{(X,Y)}(\tau)|: 0\leq \tau <n \},{\rm and} \\
\vspace{0.1cm} d_i^{{\cal B}}=\min\{ \min\limits_{X\neq Y}\{d_i^{(X,Y)} \}, \min\limits_{X \in S}\{d_i^{X}\}\}.
 \end{array}
\]
where $\Delta_i(\overrightarrow{D_{(X,Y)}}(\tau))=\emptyset$ if $D_{(X,Y)}(\tau)=\emptyset$ or $i> |D_{(X,Y)}(\tau)|$. If $X=Y$, then $D_{(X,X)}(\tau), \overrightarrow{D_{(X,X)}}(\tau)$ and $d_i^{(X,X)}$ are the same as $D(\tau),\overrightarrow{D}(\tau)$ and $d^X_i$, respectively.
Note that $D_{(Y,X)}(-\tau)\equiv D_{(X,Y)}(\tau)+\tau \pmod n$ and $\min \{g: g\in \{n\}\cup \Delta_i(\overrightarrow{D_{(X,Y)}}(\tau))\}=\min \{g: g\in \{n\}\cup \Delta_i(\overrightarrow{D_{(Y,X)}}(-\tau))\}$.

We illustrate the definition of $d^{\cal B}_i$ of a set of CDPs in the following example.
\begin{exmp}\label{24FHS set} In this example, we construct a partition-type $(24,\{\{2,3\},\{2,3\},\{2,3\}\},3)$-BNCDP,  ${\cal B}$, with $d_i^{{\cal B}}=8i$ for $1\leq i\leq 3$.

Set
{\small
\[
\begin{array}{l}
 B_{3j}^j=\{4,8\}, \ \ \ \ \ \ \ \ B_{1+3j}^j=\{0,20\}, \ \ \ \ \ B_{2+3j}^j=\{12,16\}, \\
 B_{3+3j}^j=\{9,22,23\},  B_{4+3j}^j=\{1,14,15\},  B_{5+3j}^j=\{6,7,17\}, \\
 B_{6+3j}^j=\{2,19,21\},  B_{7+3j}^j=\{11,13,18\},  B_{8+3j}^j=\{3,5,10\},\\
 \end{array}
\]
}
\noindent where $j\in \{0,1,2\}$ and the addition in the subscript is performed modulo 9.
Let ${\cal B}^j=\{B_r^j: 0\leq r\leq 8\}$ for $j\in \{0,1,2\}$ and ${\cal B}=\{{\cal B}^0, {\cal B}^1, {\cal B}^2\}$.

Clearly,
{\small
\[
\begin{array}{l}
\Delta({\cal B}^j)=3\cdot[\{\pm1,\pm2,\pm4,\pm5,\pm10,\pm11,\pm17\}],\\
\Delta_E({\cal B}^0,{\cal B}^1)=\Delta_E({\cal B}^0,{\cal B}^2)=\Delta_E({\cal B}^1,{\cal B}^2)=3\cdot [\mathbb{Z}_{24}\setminus \{0,8,16\}].\\
\end{array}
\]
}
Then ${\cal B}$ is a partition-type $(24,\{\{2,3\},\{2,3\},\{2,3\}\},3)$-BNCDP.

For each $\tau\in \Delta({\cal B}^j)$, $0\leq j<3$, the orbit cycle $\overrightarrow{D_{({\cal B}^j,{\cal B}^j})}(\tau)$ of $\tau$ is of the form
$(a,a+8,a+16)$ for some $a$, for example,
$\overrightarrow{D_{({\cal B}^j,{\cal B}^j)}}(4)=(4,12,20)$. For $\tau' \in \Delta_E({\cal B}^k,{\cal B}^j)$, $0\leq k<j\leq 2$,
the orbit cycle $\overrightarrow{D_{({\cal B}^k,{\cal B}^j})}(\tau')$ of $\tau'$ is also of the form
$(a,a+8,a+16)$ for some $a$, for example, $\overrightarrow{D_{({\cal B}^0,{\cal B}^1)}}(22)=(4,12,20)$.
Therefore, $\Delta_i(\overrightarrow{D_{({\cal B}^j,{\cal B}^j)}}(\tau))=\Delta_i(\overrightarrow{D_{({\cal B}^k,{\cal B}^j)}}(\tau'))=8i$ for $1\leq i\leq 3$.
So, for $0\leq k,j\leq 2$ and  $1\leq i\leq 3$, we have
\[
\begin{array}{l}
d_i^{({\cal B}^k,{\cal B}^j)}=\min\limits_{0\leq \tau<24}\min \{g: g\in \{24\}\cup\Delta_i(\overrightarrow{D_{({\cal B}^k,{\cal B}^j)}}(\tau))\}=8i, \\
d_i^{{\cal B}}=\min\{ \min\limits_{0\leq k<j\leq 2}\{d_i^{({\cal B}^k,{\cal B}^j)}\}, \min\limits_{0\leq j\leq 2}\{d_i^{({\cal B}^j,{\cal B}^j)}\}\}=8i.
\end{array}
\]
By direct check or by Theorem \ref{optimal sets character}, the corresponding FHS set is a strictly optimal $(24,3,3;9)$-FHS set with respect to the bound $(\ref{Bound 6})$.
\end{exmp}


The following theorem clarifies the connection between strictly optimal FHS sets and partition-type BNCDPs with a special property.

\begin{theorem}\label{optimal sets character}
There is a strictly optimal $(n,M,\lambda;l)$-FHS set $S$ with respect to the bound $(\ref{Bound 6})$ if and only if there exists a partition-type $(n,\{K_X:X\in S\},\lambda)$-BNCDP of size $l$, ${\cal B}=\{{\cal B}_X=\{B_0^X,B_1^X,\ldots,B_{l-1}^X\}:~ X\in S\}$  over $\mathbb{Z}_n$ with $d_i^{{\cal B}} \geq\left\lfloor \frac{ni}{\lambda} \right \rfloor$ for $1\leq i\leq \lambda$, where
$K_X=\{|B_r^X|:~0\leq r\leq l-1\}$, $\lambda= \left\lceil\frac{2InM-(I+1)Il}{(nM-1)M}\right\rceil$ and $I= \left\lfloor\frac{nM}{l}\right\rfloor$.
\end{theorem}

\begin{IEEEproof}
Let $S$ be a set of $M$ sequences of length $n$ over $F=\mathbf{I}_l$. Let $B_r^X=\{t:~ x(t)=r, 0\leq t\leq n-1\}$ for each $r \in F$ and $X\in S$. Then $B_r^X$ is the set of position indices in the sequence $X=(x(0),\ldots,x(n-1))$ at which the frequency $r$ appears. By Theorem \ref{FHS set=DPs}, the set $S$ of $M$ sequences is an $(n,M,\lambda;l)$-FHS set over $F$ if and only if $\{{\cal B}_X=\{B_0^X,\ldots,B_{l-1}^X\}:~ X\in S\}$ is a partition-type $(n,\{K_X:X\in S\},\lambda)$-BNCDP, where $K_X=\{|B_r^X|:~0\leq r\leq l-1\}$. It is left to show that $S$ is strictly optimal with respect to the bound $(\ref{Bound 6})$ if and only if $d_i^{{\cal B}} \geq\left\lfloor \frac{ni}{\lambda} \right \rfloor$ for $1\leq i\leq \lambda$, where ${\cal B}=\{{\cal B}_X:~X\in S\}$.

By Lemma \ref{optimal set}, $S$ is strictly optimal with respect to the bound $(\ref{Bound 6})$ if and only if $H(S;L)\leq \left\lceil\frac{\lambda L}{n} \right\rceil$ for $1\leq L\leq n$, i.e., for $1\leq i\leq \lambda$,  the inequality $H(S;L)\leq i$ holds for $\left\lfloor\frac{n(i-1)}{\lambda}\right\rfloor <L \leq \left\lfloor\frac{ni}{\lambda}\right\rfloor$. We show that  the inequality $H(S;L)\leq i$ holds for $\left\lfloor\frac{n(i-1)}{\lambda}\right\rfloor <L \leq \left\lfloor\frac{ni}{\lambda}\right\rfloor$ if and only if $d_i^{{\cal B}}\geq \left\lfloor \frac{ni}{\lambda}\right \rfloor$.

By the definition of $H_{X,Y}(\tau;j|L)$ and $D_{(X,Y)}(\tau)$, we have $H_{X,Y}(\tau;j|L)=|D_{(X,Y)}(\tau;j|L)|$, where $0<\tau <n$ if $X=Y$,  $0\leq \tau <n$ if $X\neq Y$, $D_{(X,Y)}(\tau;j|L)=D_{(X,Y)}(\tau)\cap \{j,j+1,\ldots, j+L-1\}$ and the arithmetic $j+t$ $(1\leq t\leq L-1)$ is reduced modulo $n$. Denote $d_i^{(X,Y)}(\tau)=\min \{g: g\in \{n\}\cup \Delta_i(\overrightarrow{D_{(X,Y)}}(\tau))\}$.
Then, $\max\limits_{0\leq j < n}\{|D_{(X,Y)}(\tau;j|L)|\}\leq i$ if and only $L\leq d_i^{(X,Y)}(\tau)$.
Hence, $H(X,Y;L)=\max\limits_{\tau }\max\limits_{0\leq j < n}\{|D_{(X,Y)}(\tau;j|L)|\}\leq i$ if and only if $L\leq d_i^{(X,Y)}(\tau)$ for $0\leq \tau<n$ if $X\neq Y$ and for $1\leq \tau<n$ if $X=Y$, i.e., $L\leq d_i^{(X,Y)}$=$ \min\limits_{\tau}\{d_i^{(X,Y)}(\tau)\}$. It follows that $H(X,Y;L)\leq i$ for $\left\lfloor\frac{n(i-1)}{\lambda}\right\rfloor <L \leq \left\lfloor\frac{ni}{\lambda}\right\rfloor$ if and only if
$d_i^{(X,Y)}\geq \left\lfloor \frac{ni}{\lambda}\right \rfloor$.
Hence, $H(S;L)=\max\limits_{0\leq j < n}\{\max\limits_{1\leq \tau < n,\atop X\in S}\{H_{X,X}(\tau;j|L)\},\max\limits_{0\leq \tau < n,\atop X\neq Y}\{H_{X,Y}(\tau;j|L)\}\}\leq i$ for $\left\lfloor\frac{n(i-1)}{\lambda}\right\rfloor <L \leq \left\lfloor\frac{ni}{\lambda}\right\rfloor$ if and only if $d_i^{(X,Y)}\geq \left\lfloor \frac{ni}{\lambda} \right \rfloor$, for $X,Y\in S$, which is equivalent to $d_i^{{\cal B}}=\min\{d_i^{(X,Y)}:X,Y\in S\}\geq \left\lfloor \frac{ni}{\lambda}\right \rfloor$.
This completes the proof.
\end{IEEEproof}

Let ${\cal B}_j$ be an $(mg,g,K_j,1)$-CRDP over $\mathbb{Z}_{mg}$ for $0\leq j< M$, where ${\cal B}_j=\{B_0^j,B_1^j,\ldots,B_{u-1}^j\}$. The set $\{{\cal B}_0,\ldots,{\cal B}_{M-1}\}$ is referred to as an $(mg,g,\{K_0,K_1,\ldots,K_{M-1}\},1)$-BNCRDP ({\rm balanced nested cyclic relative difference packing}) over $\mathbb{Z}_{mg}$, if $\Delta({\cal B}_j,{\cal B}_{j'})$ contains each element of $\mathbb{Z}_{mg}\setminus m\mathbb{Z}_{mg}$ at most once and no element of $m\mathbb{Z}_{g}$ occurs for $0\leq j\neq j'<M$. For convenient, the number $u$ of the base blocks in ${\cal B}_j$ is also said to be the size of the BNCDP.

One importance of a BNCRDP is that we can put an appropriate BNCDP on its subgroup to derive a new BNCDP.

\begin{lemma}\label{CRDPs=>CDPs}
Suppose that there exists an $(mg,g,\{K_0,\ldots,K_{M-1}\},1)$-BNCRDP of size $u$, $\{{\cal B}_0,\ldots,{\cal B}_{M-1}\}$, such that all elements of base blocks of ${\cal B}_j$, together with $0,m,\ldots,(s-1)m$, form a complete system of representatives for the cosets of $sm\mathbb{Z}_{mg}$ in $\mathbb{Z}_{mg}$ for $0\leq j < M$, where $s|g$. If there exists a partition-type $(g,\{K_0',\ldots,K_{M-1}'\},\frac{g}{s})$-BNCDP of size $l$, ${\cal A}$ such that $d_i^{\cal A}\geq si$ for $1\leq i\leq \frac{g}{s}$. Then there exists a partition-type $(mg,\{K_0\cup K_0',\ldots,K_{M-1}\cup K_{M-1}'\},\frac{g}{s})$-BNCDP of size $\frac{gu}{s}+l$, ${\cal D}$ such that $d_i^{\cal D} \geq ism$ for $1\leq i\leq \frac{g}{s}$.
\end{lemma}

\begin{IEEEproof}
Denote ${\cal B}_j=\{B_k^j:~0\leq k<u\}$ for $0\leq j<M$ and let ${\cal A}=\{{\cal A}_0, {\cal A}_1,\ldots, {\cal A}_{M-1}\}$ be a partition-type $(g,\{K_0',\ldots,K_{M-1}'\},\frac{g}{s})$-BNCDP of size $l$ over $\mathbb{Z}_{g}$ with $d_i^{\cal A}\geq si$ for $1\leq i\leq \frac{g}{s}$, where $ {\cal A}_j=\{A_r^j:0\leq r<l\}$.  For $0\leq j <M$, set

\[
\begin{array}{l}
\vspace{0.2cm}{\cal A}'_j=\{mA_r^j:~0\leq r<l\}, \\
\vspace{0.2cm} B_{(k,i')}^j=B_k^j+i'sm\ {\rm for}\ 0\leq k<u \ {\rm and}\ 0\leq i'<\frac{g}{s}, \ {\rm and}\\
\vspace{0.2cm}{\cal D}_j=\{B_{(k,i)}^j:~0\leq k<u, 0\leq i<\frac{g}{s} \}\cup {\cal A}'_j,
\end{array}
\]
then the size of ${\cal D}_j$ is $\frac{gu}{s}+l$.

Let ${\cal D}=\{{\cal D}_j:0\leq j<M\}$. It remains to prove that ${\cal D}$ is a partition-type $(mg,\{K_0\cup K_0',\ldots,K_{M-1}\cup K_{M-1}'\},\frac{g}{s})$-BNCDP over $\mathbb{Z}_{mg}$ with $d_i^{\cal D}\geq ism$ for $1\leq i\leq \frac{g}{s}$.

Since all elements of base blocks of ${\cal B}_{j}$, together with $0,m,\ldots,(s-1)m$, form a complete system of representatives for the cosets of $sm\mathbb{Z}_{mg}$ in $\mathbb{Z}_{mg}$, we have $\{B_{(k,i)}^j:~0\leq k<u, 0\leq i<\frac{g}{s} \}$ is a partition of $\mathbb{Z}_{mg}\setminus m\mathbb{Z}_{mg}$. Since ${\cal A}_j$ is a partition-type $(g,\{K_0',\ldots,K_{M-1}'\},\frac{g}{s})$-BNCDP, we have that ${\cal A}'_j$ is a partition of $m\mathbb{Z}_{mg}$ and ${\cal D}_j$ is a partition of $\mathbb{Z}_{mg}$.

For $0\leq j\neq j'<M$,
\[
\begin{array}{l}
\vspace{0.2cm}\Delta_E({\cal D}_j,{\cal D}_{j'}) \\
\vspace{0.2cm}=\left(\bigcup\limits_{k=0}^{u-1} \bigcup\limits_{i'=0}^{\frac{g}{s}-1}\Delta_E(B_{(k,i')}^j,B_{(k,i')}^{j'})\right)\bigcup \Delta_E({\cal A}'_j,{\cal A}'_{j'})\\
\vspace{0.3cm}=\left(\bigcup\limits_{i'=0}^{\frac{g}{s}-1}\Delta_E({\cal B}_j+i'sm,{\cal B}_{j'}+i'sm)\right)\bigcup \Delta_E({\cal A}'_j,{\cal A}'_{j'}),\\
=\frac{g}{s}\cdot[\Delta_E({\cal B}_j,{\cal B}_{j'})]\bigcup (m\Delta_E({\cal A}_j,{\cal A}_{j'})).
\end{array}
\]
Since $\{{\cal B}_j:~0\leq j <M\}$ is an $(mg,g,\{K_0,\ldots,K_{M-1}\},1)$-BNCRDP, we have that $\Delta({\cal B}_j)\subset \mathbb{Z}_{mg}\setminus m\mathbb{Z}_{mg}$ and $\Delta_E({\cal B}_j,{\cal B}_{j'})\subset \mathbb{Z}_{mg}\setminus m\mathbb{Z}_{mg}$ for $0\leq j\neq j' <M$. Since $\{{\cal A}_j:~0\leq j\leq M-1 \}$ is a $(g,\{K_0,\ldots,K_{M-1}\},\frac{g}{s})$-BNCDP, we have that $\Delta({\cal A}'_j)=m\Delta({\cal A}_j)$ contains each non-zero element of $ m\mathbb{Z}_{mg}$ at most $\frac{g}{s}$ times and $\Delta_E({\cal A}'_j,{\cal A}'_{j'})=m\Delta({\cal A}_j,{\cal A}_{j'})$ contains each element of $ m\mathbb{Z}_{mg}$ at most $\frac{g}{s}$ times for $j\neq j'$. It follows that $\Delta_E({\cal D}_j,{\cal D}_{j'})$
contains each element of $\mathbb{Z}_{mg}$ at most $\frac{g}{s}$ times. Similarly, $\Delta({\cal D}_j)$ contains each non-zero element of $\mathbb{Z}_{mg}$ at most $\frac{g}{s}$ times. Hence, ${\cal D}$ is a partition-type $(mg,\{K_0\cup K_0',\ldots,K_{M-1}\cup K_{M-1}'\},\frac{g}{s})$-BNCDP of size $\frac{gu}{s}+l$ over $\mathbb{Z}_{mg}$.

From the construction, it is easy to see that if $\tau \in \Delta_E({\cal B}_j,{\cal B}_{j'})$, then the orbit cycle $\overrightarrow{D_{(j,j')}}(\tau)$ of $\tau$ in $({\cal D}_j,{\cal D}_{j'})$ is of the form $(a_0,a_0+sm,a_0+2sm,\ldots, a_0+mg-sm)$, and if $\tau\in \Delta_E({\cal A}'_j,{\cal A}'_{j'})$ then the orbit cycle $\overrightarrow{D_{(j,j')}}(\tau)$ of $\tau$ in $({\cal D}_j,{\cal D}_{j'})$  is of the form $(mb_0,mb_1,\ldots,mb_{s'-1})$ where $(b_0,b_1,\ldots,b_{s'-1})$ is the orbit cycle of $\frac{\tau}{m}$ in $({\cal A}_j,{\cal A}_{j'})$, otherwise ${D_{(j,j')}}(\tau)=\emptyset$.
Then $d_i^{({\cal D}_j,{\cal D}_{j'})}(\tau)=ism$ for $\tau \in \Delta({\cal B}_j,{\cal B}_{j'})$, $d_i^{({\cal D}_j,{\cal D}_{j'})}(\tau)\geq ism$ for $\tau\in \Delta_E({\cal A}'_j,{\cal A}'_{j'})$ because $d_i^{({\cal A}_j,{\cal A}_{j'})}\geq si$ for $1\leq i\leq \frac{g}{s}$ and  $d_i^{({\cal D}_j,{\cal D}_{j'})}(\tau)=mg$ for $\tau \notin \Delta_E({\cal D}'_j,{\cal D}'_{j'})$. It follows that $d_i^{({\cal D}_j,{\cal D}_{j'})}=\min\{d_i^{({\cal D}_j,{\cal D}_{j'})}(\tau):1\leq \tau<mg\}=ism$ for $1\leq i\leq \frac{g}{s}$. Similarly, it is readily checked that $d_i^{{\cal D}_j}=ism$ for $1\leq i\leq \frac{g}{s}$.
Therefore, $d_i^{{\cal D}}=\min\{\min\limits_{0\leq j\neq j'<M} \{d_i^{({\cal D}_j,{\cal D}_{j'})}\}, \min\limits_{0\leq j<M} \{d_i^{{\cal D}_j}\}\}=ism$ for $1\leq i \leq \frac{g}{s}$, and ${\cal D}$ is the required BNCDP. This completes the proof.
\end{IEEEproof}

\section{Combinatorial Constructions} %

\subsection{Direct constructions of strictly optimal $(n,2;\lfloor\frac{n}{2}\rfloor)$-FHSs} %
\label{length2}                                                                                   %

{\bf Construction A1}:
Let $u$ be a positive integer and let $X=\{x(t)\}_{t=0}^{2u-1}$ be the FHS of length $2u$ over $\mathbb{Z}_u$ defined by $x(t)=x(2t_1+t_0)=(-1)^{t_0}t_1$, where $t=2t_1+t_0$, $0\leq t_0 <2,\ 0\leq t_1 <u$.

\begin{theorem}\label{ strictly optimal 2u}
For a positive integer $u$, the FHS $X$ in Construction A$1$ is a strictly optimal $(2u,2;u)$-FHS, and $H(X;L)=\lceil \frac{L}{u}\rceil$ for $1\leq L\leq 2u$.
\end{theorem}

\noindent \begin{IEEEproof}
For $0\leq r < u$, denote $B_r=\{t:~ x(t)=r, 0\leq t< 2u\}$, then we have
\[
\begin{array}{l}
\vspace{0.1cm}B_{j}=\{2j,1-2j\}\pmod {2u}\  {\rm for}\ 0 \leq j < u, \\
\vspace{0.1cm}\overrightarrow{D}(1-4j)=(2j, 2j+u)\pmod {2u}\  {\rm for}\ 0 \leq j \leq\lfloor \frac{u-1}{2}\rfloor,  \\
\vspace{0.1cm}\overrightarrow{D}(1-4j)=(2j-u, 2j)\pmod {2u}\  {\rm for}\ \lfloor \frac{u+1}{2}\rfloor \leq j < u,  \\
\vspace{0.1cm}D(k)= \emptyset,\ {\rm if}\ k\neq \pm (1-4j), \  0 \leq j < u.
\end{array}
\]
By the fact $\min \{g: g\in \{2u\}\cup \Delta_{i}(\overrightarrow{D}(\tau))\}=\min \{g: g\in \{2u\}\cup \Delta_{i}(\overrightarrow{D}(-\tau))\}$, we have $d_{1}=u$ and $d_{2}=2u$. Then, $\{B_0,\cdots, B_{u-1}\}$ is a partition-type $(2u,2,2)$-CDP of size $u$ with $d_{i}=iu=\left\lfloor \frac{2ui}{\lfloor \frac{2u}{u} \rfloor} \right\rfloor$ for $0<i\leq 2$. Therefore, by Theorem {\ref{optimal character}}, $X$ is strictly optimal with respect to the bound ({\ref{Bound 4}}).
\end{IEEEproof}

Remark: When $u$ is an odd integer, a strictly optimal $(2u,2;u)$-FHS can be also obtained from \cite{CZYT2014}.

{\bf Construction A2}:
Let $n$ be an odd integer and let $X=\{x(t)\}_{t=0}^{n-1}$ be the FHS of length $n$ over $\mathbb{Z}_{(n-1)/2}$ defined by $x(t)=j$ for $t \in B_j$, $0\leq j< \frac{n-1}{2}$, where $B_j$ are defined as follows:

\noindent If $n=8a+1$, then
\[
\begin{array}{l}
\vspace{0.1cm}B_0=\{0, 4a+1, 8a\};\ \
\vspace{0.1cm}B_1=\{4a-1, 4a\};\ \ \\
\vspace{0.1cm}B_{1+r}=\{r, 2a-2+2r\}\ {\rm for}\ 1\leq r \leq a;\ \ \\
\vspace{0.1cm}B_{a+1+r}=\{a+r, 2a-1+2r\}\ {\rm for}\ 1\leq r \leq a-1; \\
\vspace{0.1cm}B_{2a+r}=B_{1+r}+4a+1\pmod n\ {\rm for}\ 1\leq r \leq 2a-1.
\end{array}
\]
If $n=8a+3$, then
\[
\begin{array}{l}
\vspace{0.1cm}B_0=\{0, 4a+1, 4a+2\};\ \
B_1=\{2a, 6a+3\};\ \ \\
\vspace{0.1cm}B_2=\{2a+1, 6a+1\};\ \
B_3=\{6a+2, 6a+4\}; \\
\vspace{0.1cm}B_{3+r}=\{r, 4a+1-r\} \ {\rm for}\ 1\leq r \leq 2a-1;\ \ \\
\vspace{0.1cm}B_{2a+2+r}=B_{3+r}+4a+2\pmod n\  {\rm for}\ 1\leq r \leq 2a-2.
\end{array}
\]
If $n=8a+5$, then
\[
\begin{array}{l}
\vspace{0.1cm}B_0=\{0, 4a+2, 4a+3\};\ \
B_1=\{2a+1, 6a+5\};\ \ \\
\vspace{0.1cm}B_2=\{2a+2, 6a+3\};\ \
B_3=\{1, 6a+4\}; \\
\vspace{0.1cm}B_{3+r}=\{2a+2+r, 6a+3-r\} \ {\rm for}\ 1\leq r \leq 2a-1;\ \ \\
\vspace{0.1cm}B_{2a+2+r}=B_{3+r}+4a+3\pmod n\  {\rm for}\ 1\leq r \leq 2a-1.
\end{array}
\]
If $n=8a+7$, then
\[
\begin{array}{l}
\vspace{0.1cm}B_0=\{0, 4a+3, 4a+4\};\ \\
\vspace{0.1cm}B_{r}=\{r, 2a+2r\}\ {\rm for}\ 1\leq r \leq a+1;\ \ \\
\vspace{0.1cm}B_{a+r}=\{a+1+r, 2a+1+2r\}\ {\rm for}\ 1\leq r \leq a; \\
\vspace{0.1cm}B_{2a+r}=B_{r}+4a+4\pmod n \ {\rm for}\ 1\leq r \leq 2a+1.
\end{array}
\]

\begin{theorem}\label{n}
For any odd integer $n\geq 5$, the FHS $X$ in Construction A$2$ is a strictly optimal $(n,2;\frac{n-1}{2})$-FHS and $H(X;L)=\lceil \frac{2L}{n}\rceil$ for $1\leq L\leq n $.
\end{theorem}
\begin{IEEEproof}
By Theorem \ref{optimal character}, we only need to show that $\{B_j:~0\leq j < \frac{n-1}{2}\}$ is a partition-type $(n,\{2,3\},2)$-CDP with $d_{i}\geq \left\lfloor\frac{ni}{\lfloor\frac{n}{(n-1)/2}\rfloor}\right \rfloor=\lfloor\frac{ni}{2}\rfloor$ for $1\leq i \leq \lfloor\frac{n}{(n-1)/2}\rfloor=2$.
Clearly,  ${\cal B}=\{B_j:~0\leq j < \frac{n-1}{2}\}$ is a partition of $\mathbb{Z}_n$. By the fact $\min \{g: g\in \{n\} \cup \Delta_{i}(\overrightarrow{D}(\tau))\}=\min \{g: g\in \{n\} \cup \Delta_{i}(\overrightarrow{D}(-\tau))\}$, we only need to compute $\overrightarrow{D}(\tau)$ for $1\leq \tau \leq \frac{n-1}{2}$, so as to obtain $d_{i}$. According to the construction, we divide it into four cases.

Case 1: $n=8a+1$. By Construction A$2$, we have that
\[
\begin{array}{l}
\vspace{0.1cm}\overrightarrow{D}(1)=(4a-1, 8a),   \vspace{0.1cm}\overrightarrow{D}(4a)=(1+4a), \\
\vspace{0.1cm}\overrightarrow{D}(4a-1)= (1+4a), \\
\vspace{0.1cm}\overrightarrow{D}(2a-2+r)=(r,r+4a+1),   1\leq r\leq a, \ {\rm and} \\
\vspace{0.1cm}\overrightarrow{D}(a-1+r)= (a+r,5a+1+r), 1\leq r\leq a-1.
\end{array}
\]
Then, it holds that $d_{1}=4a$ and $\lambda=\max\{|D(\tau)|:~0< \tau <4a\}=2$. Therefore, by Theorem {\ref{optimal character}}, the sequence $X$ is strictly optimal with respect to the bound ({\ref{Bound 4}}).

Case 2: $n=8a+3$. By Construction A$2$, we have
\[
\begin{array}{l}
\vspace{0.1cm}\overrightarrow{D}(1)=(4a+1),   \vspace{0.1cm}\overrightarrow{D}(4a+1)=(0,2+4a), \vspace{0.1cm}\overrightarrow{D}(2)=(2+6a), \\
\vspace{0.1cm}\overrightarrow{D}(4a)=(1+2a,3+6a),\vspace{0.1cm}\overrightarrow{D}(3)=(2a-1), \ {\rm and}\\
\vspace{0.1cm}\overrightarrow{D}(4a+1-2r)=(r,r+4a+2),   1\leq r\leq 2a-2.
\end{array}
\]
Then, it holds that $d_{1}=4a+1$ and $\lambda=\max\{|D(\tau)|:~0< \tau <4a+1\}=2$. Therefore, by Theorem {\ref{optimal character}}, the sequence $X$ is strictly optimal with respect to the bound ({\ref{Bound 4}}).

Case 3: $n=8a+5$. By Construction A$2$, we have
\[
\begin{array}{l}
\vspace{0.1cm}\overrightarrow{D}(1)=(4a+2),   \vspace{0.1cm}\overrightarrow{D}(4a+2)=(0,3+4a), \\
\vspace{0.1cm}\overrightarrow{D}(2a+2)=(4+6a), \vspace{0.1cm}\overrightarrow{D}(4a+1)= (2a+2,5+6a), \ {\rm and}\\
\vspace{0.1cm}\overrightarrow{D}(4a+1-2r)=(2a+2+r,r+6a+5),   1\leq r\leq 2a-1.
\end{array}
\]
Then, it holds that $d_{1}=4t+2$ and $\lambda=\max\{|D(\tau)|:~0< \tau <4a+2\}=2$. Therefore, by Theorem {\ref{optimal character}}, the sequence $X$ is strictly optimal with respect to the bound ({\ref{Bound 4}}).

Case 4: $n=8t+7$. By Construction A$2$, we have
\[
\begin{array}{l}
\vspace{0.1cm}\overrightarrow{D}(1)=(4t+3),   \vspace{0.1cm}\overrightarrow{D}(4t+3)=(0,4+4t), \\
\vspace{0.1cm}\overrightarrow{D}(2t+r)=(r,r+4t+4)\ {\rm for}\ 1\leq r\leq t+1, \ {\rm and}\\
\vspace{0.1cm}\overrightarrow{D}(t+r)=(t+1+r,r+5t+5),\  1\leq r\leq t.
\end{array}
\]
Then, it holds that $d_{1}=4a+3$ and $\lambda=\max\{|D(\tau)|:~0< \tau <4t+3\}=2$.  Therefore, by Theorem {\ref{optimal character}}, the sequence $X$ is strictly optimal with respect to the bound ({\ref{Bound 4}}).
This completes the proof.
\end{IEEEproof}

\subsection{Strictly optimal FHSs from resolvable CRDFs}

In this subsection, we construct resolvable CRDFs by using cyclotomic classes in order to obtain strictly optimal FHSs.



Resolvable CRDFs have been intensively studied in design theory, see \cite{CD2007}.
We quote some results of resolvable CRDFs in the following lemma.
\begin{lemma}\label{Result-CRDF}
(1) There exists a resolvable $(v,2,3,1)$-CRDF for all
$v$ of the form $2^kp_1p_2\cdots p_s$ where $k\in \{1,5\}$ and each $p_j \equiv 1 \pmod {12}$ is a prime (\cite{BZ2001}, {\rm \cite{BZ1998}}).

(2) There exists a resolvable $(v,2,3,1)$-CRDF for all
$v$ of the form $8p_1p_2\cdots p_s$ where each $p_j \equiv 1 \pmod {6}$ is a prime (\cite{BZ2001}).


(3) There exists a resolvable $(v,3,4,1)$-CRDF for all
$v$ of the form $3p_1p_2\cdots p_s$ where each $p_j \equiv 1 \pmod {4}$ is a prime ({\rm \cite{BZ1998}}).
\end{lemma}

The application of Lemma \ref{CRDP=>CDP} to the CRDFs in Lemma \ref{Result-CRDF} yields the following strictly optimal FHSs.
\begin{theorem}\label{ A FHS-CRDF}
(1) There exists a strictly optimal $(v, 2; \frac{v+1}{3})$-FHS for all
$v$ of the form $2^kp_1p_2\cdots p_s$ where $k\in \{1,5\}$ and each $p_i\equiv 1 \pmod {12}$ is a prime.

(2) There exists a strictly optimal $(v, 2; \frac{v+1}{3})$-FHS for all
$v$ of the form $8p_1p_2\cdots p_s$ where each $p_i\equiv 1 \pmod {6}$ is a prime.


(3) There exists a strictly optimal $(v, 3; \frac{v+1}{4})$-FHS for all
$v$ of the form $3p_1p_2\cdots p_s$ where each $p_i$ $\equiv 1 \pmod {4}$ is a prime.
\end{theorem}


Let $q$ be a prime power with $q=ef+1$ and $GF(q)$ be the finite field of $q$ elements. Given a primitive element $\alpha$ of $GF(q)$, define $C_0^{e}=\{\alpha^{je}:0\leq j\leq f-1\}$, the multiplicative group generated by $\alpha^e$, and $$C_i^{e}=\alpha^iC_0^{e}$$
for $1\leq i\leq e-1$. Then $C_0^{e},  C_1^{e},\ldots,  C_{e-1}^{e}$ partition $GF(q)^*=GF(q)\setminus \{0\}$. The $C_i^{e}$ $(0\leq i<e)$ are known as {\em cyclotomic classes} of index $e$ (with respect to $GF(q)$). Given a list $\{a_1,a_2,\ldots,a_e\}$ of elements in $GF(q)^*$, if each cyclotomic class $C_i^{e}$, $0\leq i\leq e$, contains exactly one element of the list, then we say that the list forms a {\em complete system of distinct representative} of cyclotomic classes.

In the theory of cyclotomy, the numbers of solutions of
$$x+1=y,\ x\in C_i^{e},\ y\in C_j^{e}$$
are called {\em cyclotomic numbers} of order $e$ respect to $GF(q)$ and denoted by $(i,j)_e$.

\begin{lemma}\label{4-CRDF}
There exists a resolvable $(4p,4,3,1)$-CRDF for any prime $p\equiv 7 \pmod {12}$.
\end{lemma}
\begin{IEEEproof}
Let $\varepsilon$ be a primitive sixth root of unity in $\mathbb{Z}_p$. Clearly, $\varepsilon^5$ is also a primitive sixth root of unity and $1-\varepsilon+\varepsilon^2=0$. Thus, $\frac{\varepsilon^5+1}{\varepsilon+1}=\frac{-\varepsilon^2+1}{\varepsilon+1}=1-\varepsilon=-\varepsilon^2\in C_1^2$ since $-1 \in C_1^2$. Without loss of generality, we may assume that $\varepsilon+1 \in C_0^2$. Since $gcd(4,p)=1$, we have that $\mathbb{Z}_{4p}$ is isomorphic to $\mathbb{Z}_4\times \mathbb{Z}_p$.  Denote by $R$ a complete system of representatives for the cosets of $\{1,\varepsilon^2,\varepsilon^4\}$ in $C_0^2$, this implies that
\begin{equation}\label {C_0^2}
\{r \varepsilon^{2j}:~r\in R, 0\leq j<3\}=C_0^2.
\end{equation}
 Set
\[
\begin{array}{l}
 \vspace{0.2cm} {\cal B}=\{ \{(0,r),(0,r\varepsilon^2),(0,r\varepsilon^4)\}:~r \in R \}  \\
 \vspace{0.2cm}\hspace{0.8cm} \bigcup \{ \{(1,-r\varepsilon^2),(2,-r),(3,-r\varepsilon)\}:~r \in R\}\\
 \vspace{0.2cm}\hspace{0.8cm} \bigcup\{ \{(1,-r\varepsilon^4),(2,-r\varepsilon^2),(3,-r\varepsilon^3)\}:~r \in R \}  \\
 \vspace{0.2cm}\hspace{0.8cm} \bigcup \{  \{(1,-r),(2,-r\varepsilon^4),(3,-r\varepsilon^5)\}:~r \in R\}.
\end{array}
\]
Then  $\bigcup\limits_{B\in \cal B}B$
\[
\begin{array}{l}\\
 \vspace{0.2cm} \equiv (\bigcup\limits_{r \in R } \{1\} \times \{-r,-r\varepsilon,-r\varepsilon^2,-r\varepsilon^3,-r\varepsilon^4,-r\varepsilon^5\}) \\
 \vspace{0.2cm}\hspace{0.4cm} \bigcup(\bigcup\limits_{r \in R } \{0\} \times \{r,r\varepsilon^2,r\varepsilon^4,-r,-r\varepsilon^2,-r\varepsilon^4\}) \\
 \hspace{4cm} \pmod{\{(0,0),(2,0)\}}.
\end{array}
\]
In view of (\ref{C_0^2}), $s=\frac{4}{3-1}=2$ and $-1, \varepsilon \in C_1^2$, we have
\[
\begin{array}{l}
 \vspace{0.2cm} \bigcup\limits_{B\in \cal B}B \cup \{(0,0),(1,0)\}\\
 \equiv \bigcup\limits_{i=0}^{1}\{i\}\times \mathbb{Z}_p \pmod{\{(0,0),(2,0)\}}. \\
\end{array}
\]
Hence,  all elements of base blocks of ${\cal B}$, together with $(0,0)$ and $(1,0)$, form a complete system of representatives for the cosets $\{(0,0),(2,0)\}$ in $\mathbb{Z}_4\times \mathbb{Z}_p$. It is left to check that ${\cal B}$ is a $(4p,4,3,1)$-CRDF.

It is straightforward that
\[
\begin{array}{l}
 \vspace{0.2cm} \Delta({\cal B})=\bigcup \limits_{z=0}^3\{z\} \times \Delta_z, \\
 {\rm where}\   \Delta_0= \bigcup \limits_{r \in R}\bigcup\limits_{j=0}^2 \{ \pm r\varepsilon^{2j}(1-\varepsilon^2) \}, \\
 \Delta_1= \bigcup\limits_{r\in R} \bigcup\limits_{j=0}^{2}\{r\varepsilon^{2j}(\varepsilon^2-1), r\varepsilon^{2j}(1-\varepsilon)\}, \\
 \Delta_2= \bigcup\limits_{r\in R} \bigcup\limits_{j=0}^{2}\{\pm r\varepsilon^{2j}(\varepsilon^2-\varepsilon)\}, \\
 \Delta_3= -\Delta_1.
 \end{array}
\]
In view of (\ref{C_0^2}), $-1\in C_1^2$ and $\varepsilon+1 \in C_0^2$, then we have:
\[
\begin{array}{l}
\Delta({\cal B})=\mathbb{Z}_4 \times (\mathbb{Z}_p\setminus \{0\}).
\end{array}
\]
Therefore, ${\cal B}$ is a resolvable $(4p,4,3,1)$-CRDF.
\end{IEEEproof}

\begin{lemma}\label{6-CRDF}
There exists a resolvable $(6p,6,3,1)$-CRDF for any prime $p\equiv 5 \pmod {8}$.
\end{lemma}
\begin{IEEEproof}
For $p=5$, it exists by Example \ref{30}. For other prime $p\equiv 5 \pmod 8$, let $\varepsilon$ be a primitive fourth root of unity in $\mathbb{Z}_p$, then $\varepsilon \in C_1^2$. Let $\varepsilon \in C_a^4$ and $1+\varepsilon \in C_{2-j}^4$, where $a\in \{1,3\}$ and $j\in \{0,1,2,3\}$. We first show that there exists an element $z$ such that $\frac{2\varepsilon}{1+\varepsilon}(z+1),\frac{2\varepsilon}{1+\varepsilon}(z\varepsilon)\in C_2^4$.
For this purpose, we need the cyclotomic numbers $(j-a,j)_4$. For any prime $p\equiv 5 \pmod8$, they are given in page 48 of \cite{S1967}
\[
\begin{array}{l}
\vspace{0.2cm}(0,1)_4=(3,2)_4=\frac{p+1+2x-8y}{16},\\
\vspace{0.2cm}(0,3)_4=(1,2)_4=\frac{p+1+2x+8y}{16}, \ {\rm and}\\
\vspace{0.2cm}(1,0)_4=(2,1)_4=(2,3)_4=(3,0)_4=\frac{p-3-2x}{16},
\end{array}
\]
where $p=x^2+4y^2$ with $x\equiv 1 \pmod 4$.
Since $p=x^2+4y^2$, we have that $|x|\leq \sqrt{p}$ and $(x\pm4y)^2\leq 5(x^2+4y^2)=5p$. Thus $\frac{p+1+2x\pm 8y}{16}\geq \frac{p+1-2\sqrt{5p}}{16}> 0$ and $\frac{p-3-2x}{16}\geq \frac{p-3-2\sqrt{p}}{16}> 0$ for $p\geq 29$, i.e., for any $0\leq j < 4$, the cyclotomic number $(j-a,j)_4> 0$  if $p\geq 29$. Hence, there exists an element $z_j\in C_{j-a}^4$  such that $z_j+1\in C_{j}^4$. Thus, for each $p\equiv 5 \pmod {8}$ with $p\geq 29$ there exists an element $z$ such that $\frac{2\varepsilon}{1+\varepsilon}(z+1),\frac{2\varepsilon}{1+\varepsilon}(z\varepsilon)\in C_2^4$.
For $p=13$, take $\varepsilon=8$ and $z=11$. Then $\frac{2\varepsilon}{1+\varepsilon}(z+1),\frac{2\varepsilon}{1+\varepsilon}(z\varepsilon)\in C_2^4.$

Since $gcd(6,p)=1$, we have that $\mathbb{Z}_{6p}$ is isomorphic to $\mathbb{Z}_6\times \mathbb{Z}_p$.
Set
\[
\begin{array}{l}
 \vspace{0.2cm} {\cal B}=\{ \{(0,w),(0,-w),(1,w\varepsilon)\}:~w\in C_0^4\} \\
 \vspace{0.2cm}\hspace{0.8cm}\bigcup \{\{(1,-w),(3,w\varepsilon),(3,-w\varepsilon)\}:~w\in C_0^4\}\\
 \vspace{0.2cm}\hspace{0.cm} \bigcup \{\{(2,-w\varepsilon(2z+1)),(4,-w\varepsilon),(5,w\varepsilon(2z+1))\}:~w\in C_0^4\}\\
 \vspace{0.2cm}\hspace{0.8cm} \bigcup \{\{(2,w(2z+1)),(4,w),(5,-w(2z+1))\}:~w\in C_0^4\}.
\end{array}
\]
Then
{  \[
\begin{array}{l}
 \vspace{0.2cm} \bigcup\limits_{B\in \cal B}B\equiv \vspace{0cm}  \bigcup\limits_{w\in C_0^4 } \{0,1\} \times \{ w,-w,w\varepsilon,-w\varepsilon\} \\
 \vspace{0.2cm} \hspace{1.3cm}  \bigcup ( \bigcup\limits_{w\in C_0^4 } \{2\} \times \{w(2z+1),-w(2z+1), \\   \hspace{1.2cm} w(2z+1)\varepsilon,-w(2z+1)\varepsilon\}) \pmod{\{(0,0),(3,0)\}}.
\end{array}
\]
}
Since $\varepsilon \in C_1^2$ and $-1 \in C_2^4$, we have
\[
\begin{array}{l}
 \vspace{0.2cm} (\bigcup\limits_{B\in \cal B}B)\cup \{(0,0),(1,0),(2,0)\}\\

 \equiv \bigcup\limits_{i=0}^{2}\{i\}\times \mathbb{Z}_p \pmod{\{(0,0),(3,0)\}}.
\end{array}
\]
Hence, all elements of base blocks of ${\cal B}$, together with $(0,0),(1,0)$ and $(2,0)$, form a complete system of representatives for the cosets $\{(0,0),(3,0)\}$ in $\mathbb{Z}_6\times \mathbb{Z}_p$. It is left to check that ${\cal B}$ is a $(6p,6,3,1)$-CRDF.

It is straightforward that
\[
\begin{array}{l}
 \vspace{0.2cm}\hspace{2cm} \Delta({\cal B})=\bigcup \limits_{b=0}^5\{b\} \times \Delta_b, \\
 {\rm where}\   \Delta_0= \bigcup \limits_{w\in C_0^4} \{ \pm 2w,\pm 2w\varepsilon \}, \\
 \Delta_1= \bigcup\limits_{w\in C_0^4} \{w(\varepsilon-1), w(1+\varepsilon),w\varepsilon(2z+2), -w(2z+2)\} , \\
 \Delta_2= \bigcup\limits_{w\in C_0^4} \{w(1+\varepsilon), w(1-\varepsilon) , 2wz\varepsilon, -2wz)\} , \\
 \Delta_3= \bigcup\limits_{w\in C_0^4} \{\pm 2w\varepsilon(2z+1),\pm 2w(2z+1) \}, \\
 \Delta_4=- \Delta_2,\ {\rm and}\
 \Delta_5= -\Delta_1. \\
 \end{array}
\]
Since $\varepsilon \in C_1^2$, $-1\in C_2^4$, $\varepsilon(1+\varepsilon)=\varepsilon-1$ and $\frac{(2z+2)\varepsilon}{1+\varepsilon},\frac{-2z}{1+\varepsilon}\in C_2^4$.  We have that
\[
\begin{array}{l}
 \vspace{0.2cm} \Delta({\cal B})
=\mathbb{Z}_6 \times (\mathbb{Z}_p\setminus \{0\}).
\end{array}
\]
Therefore, ${\cal B}$ is a resolvable $(6p,6,3,1)$-CRDF.
\end{IEEEproof}

The application to the CRDFs in this subsection yields the following strictly optimal FHSs.
\begin{corollary}
(1) There exists a strictly optimal $(4p, 2; \frac{4p+2}{3})$-FHS for any prime $p \equiv 7 \pmod {12}$.

(2) There exists a strictly optimal $(6p, 2; 2p+1)$-FHS for any prime $p \equiv 5 \pmod {8}$.
\end{corollary}

\begin{IEEEproof}
In the following, we only prove the case (1). The other case can be handled similarly. By Lemma \ref{4-CRDF}, there exists a resolvable $(4p,4,3,1)$-CRDF for any prime $p\equiv 7 \pmod {12}$. By Theorem \ref{ strictly optimal 2u} and Theorem \ref{optimal character}, there is a partition-type $(4,2,2)$-CDP of size $2$ over $\mathbb{Z}_{4}$ with $d_i=2i$ for $1\leq i\leq 2$. By applying Lemma \ref{CRDP=>CDP} with $s=2$, we obtain a partition-type $(4p,\{2,3\},2)$-CDP over $\mathbb{Z}_{4p}$ with $d_i\geq 2p$ for $1\leq i\leq 2$. Further, applying Theorem \ref{optimal character} yields a strictly optimal $(4p, 2; \frac{4p+2}{3})$-FHS.
\end{IEEEproof}
\subsection{A cyclotomic construction of strictly optimal FHS sets}

In this subsection, we obtain a class of partition-type BNCDPs with a special property by using cyclotomic classes, from which we obtain a new construction of strictly optimal FHS sets.

Let $v$ be an odd integer with $v>1$ and denote by $U(\mathbb{Z}_v)$ the set of all units in $\mathbb{Z}_v$. An element $g\in U(\mathbb{Z}_v)$ is called a primitive root modulo $v$ if its multiplicative order modulo $v$ is $\varphi(v)$, where $\varphi(v)$ denotes the Euler function which counts the number of positive integers less than and coprime to $v$. It is well known that for an odd prime $p$, there exists an element $g$ such that $g$ is a primitive root modulo $p^b$ for all $b\geq 1$ \cite{A1976}.

Let $v$ be an odd integer of the form $v=p_1^{m_1}p_2^{m_2}\cdots p_s^{m_s}$ for $s$ positive integers $m_1,m_2,\ldots,m_s$ and $s$ distinct primes $p_1,p_2,\ldots,p_s$. Let $e$ be a common factor of $p_1-1,p_2-1,\ldots,p_s-1$ and $e>1$. Define $f=\min \{\frac{p_i-1}{e}:~ 1\leq i\leq s\}$. For $1\leq i \leq s$, let $g_i$ be a primitive root modulo $p_i^{m_i}$. By the Chinese Remainder Theorem, there exist unique elements $g, a\in U(\mathbb{Z}_v)$ such that
\[
\begin{array}{l}
\vspace{0.1cm} g\equiv g_i^{f_ip_i^{m_i-1}}\pmod {p_i^{m_i}}\ {\rm for\ all}\ 1\leq i\leq s, \\
a\equiv g_i \pmod {p_i^{m_i}}\ {\rm for\ all}\ 1\leq i\leq s,
\end{array}
\]
then the multiplicative order of $g$ modulo $v$ is $e$,  the list of differences arising from $G=\{1,g,\ldots, g^{e-1}\}$ is a subset of $U(\mathbb{Z}_v)$ and
$a^tg^c-g^{c'}\in U(\mathbb{Z}_v)$ for $1\leq t<f$ and $0\leq c,c'<e$.
\begin{lemma}\label{cyclotomic construction}
Let $v$ be a positive integer of the form $v=p_1^{m_1}p_2^{m_2}\cdots p_s^{m_s}$ for $s$ positive integers $m_1,m_2,\ldots,m_s$ and $s$ distinct primes $ p_1,p_2,\ldots,p_s$. Let $u$ be a positive integer such that $gcd(u,v)=1$. Let $e$ be a common factor of $u, p_1-1, p_2-1, \ldots, p_s-1$ and $e>1$, and let $f=\min \{\frac{p_i-1}{e}:~ 1\leq i\leq s\}$. Then there exists an $(uv,u,\{K_0,\ldots,K_{f-1}\},1)$-BNCRDP of size $\frac{v-1}{e}$ such that all elements of base blocks of each CRDP, together with $0$, form a complete system of representatives for the cosets of $v \mathbb{Z}_{uv}$ in $\mathbb{Z}_{uv}$ where $K_0=\cdots=K_{f-1}=\{e\}$.
\end{lemma}

\begin{IEEEproof} Let $g$ and $a$ be defined as above and denote $G=\{1,g,\ldots, g^{e-1}\}$. Then $G$ is a multiplicative cyclic subgroup of order $e$ of $\mathbb{Z}_v$ and $\Delta(G)\subset U(\mathbb{Z}_v)$. For $x,y\in \mathbb{Z}_v\setminus\{0\}$, the binary relation $\sim$ defined by $x\sim y$ if and only if there exists a $g' \in G$ such that $xg'=y$ is an equivalence relation over $\mathbb{Z}_v\setminus\{0\}$. Then its equivalence classes are the subsets $xG, x\in \mathbb{Z}_v\setminus\{0\},$ of $\mathbb{Z}_v$. Denote by $R$ a system of distinct representatives for the equivalence classes modulo $G$ of $\mathbb{Z}_v\setminus\{0\}$, then $|R|=\frac{v-1}{e}$ and

{ \begin{equation}
\label{v}
\begin{aligned}
\{ra^{b}g^j:~r\in R,\ 0\leq j< e\}=\mathbb{Z}_v\setminus\{0\},
\end{aligned}
\end{equation}
}
for any integer $b$.
Since $gcd(u,v)=1$, we have that $\mathbb{Z}_{uv}$ is isomorphic to $\mathbb{Z}_u\times \mathbb{Z}_v$.
Let
\[
\begin{array}{l}
\vspace{0.2cm} B_{r}^b=\{(\frac{ju}{e},ra^bg^j):~0\leq j < e\}\ {\rm for}\ r\in R\  {\rm and}\ 0\leq b < f, \\ {\cal B}_b=\{B_{r}^b:~ r\in R\}.
\end{array}
\]

We claim that $\{{\cal B}_b: 0\leq b < f\}$ is an $(uv,u,\{\{e\},\{e\},\ldots,\{e\}\},1)$-BNCRDP such that all elements of base blocks of ${\cal B}_t$, together with $(0,0)$, form a complete system of representatives for the cosets of $\mathbb{Z}_{u}\times \{0\}$ in $\mathbb{Z}_{u}\times \mathbb{Z}_{v}$.

In view of equality (\ref{v}), it holds that
\[
\begin{array}{l}
\vspace{0.2cm}\bigcup\limits_{r\in R}B_{r}^b \bigcup \{(0,0)\} \\
=\bigcup\limits_{r\in R}\{(\frac{ju}{e},ra^bg^j):~0\leq j < e\}\bigcup \{(0,0)\}\vspace{0.2cm} \\
\equiv \{0\}\times \mathbb{Z}_v \pmod {\mathbb{Z}_u\times \{0\}}.
\end{array}
\]
It follows that all elements of base blocks of ${\cal B}_t$, together with $(0,0)$, form a complete system of representatives for the cosets of $\mathbb{Z}_u\times \{0\}$ in $\mathbb{Z}_u\times \mathbb{Z}_v$.

For $0\leq b< f$, we show that $\Delta({\cal B}_b)$ contain each non-zero element of $\mathbb{Z}_u\times \mathbb{Z}_v$ at most once.

In view of equality (\ref{v}), we get
\[
\begin{array}{l}
\vspace{0.25cm}\hspace{0.4cm}\Delta({\cal B}_b)=\bigcup\limits_{r\in R}\Delta(\{(\frac{ju}{e},ra^bg^j):~ 0\leq j< e \})\\
\vspace{0.25cm}=\bigcup\limits_{r\in R}\{(\frac{j'u}{e},ra^bg^{j'})-(\frac{ju}{e},ra^bg^{j}):~ 0\leq j\neq j' < e \}\\
\vspace{0.25cm}=\bigcup\limits_{r\in R}\{(\frac{(j'-j)u}{e},ra^b(g^{j'}-g^{j})):~ 0\leq j\neq j' < e\}\\
\vspace{0.25cm}=\bigcup\limits_{r\in R}\{(\frac{cu}{e},ra^bg^{j}(g^c-1)):~0\leq j < e, 1\leq c < e \}\\
=(\frac{u}{e}\mathbb{Z}_u\setminus \{0\})\times(\mathbb{Z}_v\setminus \{0\}).
\end{array}
\]
Hence, each ${\cal B}_b$ is an $(uv,u,e,1)$-CRDP.

For $0\leq b\neq b'< f$,  according to equality (\ref{v}) and $a^{{b'}-b}g^c-1 \in U(\mathbb{Z}_v)$, we get
\[
\begin{array}{l}
\vspace{0.2cm}\Delta({\cal B}_b,{\cal B}_{b'})=\bigcup\limits_{r\in R}\Delta(B_{r}^b,B_{r}^{b'})\\
\vspace{0.2cm}=\bigcup\limits_{r\in R}\{(\frac{ku}{e},ra^{b'}g^k)-(\frac{ju}{e},ra^{b}g^j):~ 0\leq j,k < e \}\\
\vspace{0.2cm}=\bigcup\limits_{r\in R}\{(\frac{(k-j)u}{e},r(a^{b'}g^k-a^{b}g^j)):~ 0\leq j,k < e\}\\
\vspace{0.2cm}=\bigcup\limits_{r\in R}\{(\frac{cu}{e},ra^{b}g^j(a^{{b'}-{b}}g^c-1)):~0\leq j < e, 0\leq c < e \}\\
=(\frac{u}{e}\mathbb{Z}_u)\times(\mathbb{Z}_v\setminus \{0\}).
\end{array}
\]
Hence, $\Delta({\cal B}_b,{\cal B}_{b'})$ contains each element of $\mathbb{Z}_u\times \mathbb{Z}_v$ at most once.
This completes the proof.
\end{IEEEproof}

Start with a $(uv,u,\{K_0,\ldots,K_{f-1}\},1)$-BNCRDP in Lemma \ref{cyclotomic construction} where $K_0=\cdots=K_{f-1}=\{e\}$. Let $A_k=\{\frac{u}{e}j+k:~0\leq j<e\}$ and ${\cal A}'_b=\{A_k:~ 0\leq k <\frac{u}{e}\}$ for $0\leq b<f$, it is easy to see that $\{{\cal A}'_0,\ldots,{\cal A}'_{f-1}\}$ is a partition-type $(u,\{K_0,\ldots,K_{f-1}\},u)$-BNCDP of size $\frac{u}{e}$ with $d_i\geq i$ for $1\leq i\leq u$. Applying Lemma \ref{CRDPs=>CDPs} with $g=u$ and $s=1$ yields the following corollary.

\begin{corollary}\label{vu}
Suppose that the parameters $v, e, f, u$ are the same as those in the hypotheses of Lemma \ref{cyclotomic construction}.
Then there exists a $(uv,\{K_0,\ldots,K_{f-1}\},u)$-BNCDP of size $\frac{uv}{e}$ with $d_i\geq vi$ for $1\leq i\leq u$ where $K_0=\cdots=K_{f-1}=\{e\}$.
\end{corollary}

Furthermore, when $u=e$, by Theorem \ref{optimal sets character} the BNCDP in Corollary \ref{vu} is a strictly optimal FHS set. So, we have the following corollary.

\begin{corollary} (\cite{CZYT2014})
Suppose that the parameters $v, e, f$ are the same as those in the hypotheses of Lemma \ref{cyclotomic construction}.
Then there exists a  strictly optimal $(ev,f,e;v)$-FHS set $S$ with partial Hamming $H(S;L)=\left\lceil \frac{L}{v}\right\rceil$ for $1\leq L\leq ev$.
\end{corollary}

Lemma \ref{cyclotomic construction} interprets the generalized cyclotomic construction in \cite {CZYT2014} for FHS sets via cyclotomic cosetss. In comparison, our method is quite neat and more clear to understand.

\section{A direct construction of strictly optimal FHS sets}

In this section, we use finite fields to give a direct construction of strictly optimal FHS sets.

{\bf Construction B}\ \
Let $p$ be a prime and let $m$ be an integer with $m>1$. Let $\alpha$ be a primitive element of $GF(p^m)$ and denote $R=\{\sum_{i=1}^{m-1}a_i\alpha^i: a_i\in GF(p), 1\leq i<m\}$. Let $S=\{X^a:~ a\in R\}$ be a set of $p^{m-1}$ FHSs of length $p(p^m-1)$, where $X^a=\{X^a(t)\}_{t=0}^{p(p^m-1)-1}$ is defined by $X^a(t)=\alpha^{\langle  t\rangle_{p^m-1}}+\langle t\rangle_{p}+a$ and $\langle z \rangle_u$ denotes the least nonnegative residue of $z$ modulo $u$ for any positive integer $u$ and any integer $z$.
\begin{theorem}\label{optimal p}
Let $p$ be a prime prime and let $m$ be an integer with $m>1$. Then the FHS set in Construction B is a strictly optimal $(p(p^m-1),p^{m-1},p;p^m)$-FHS set with respect to the bound ({\ref{Bound 6}}) over the alphabet $GF(p^m)$.
\end{theorem}

\begin{IEEEproof}
Firstly, we prove that $H(S;L)\leq \left\lceil \frac{L}{p^m-1} \right\rceil$ for $1\leq L\leq p(p^m-1)$. For $0\leq\tau, j\leq p(p^m-1)-1$ and $a,b\in R$, the partial Hamming correlation $H_{X^a,X^b}(\tau)$ is given by
\[
\begin{array}{l}
\vspace{0.2cm} H_{X^a,X^b}(\tau:j|L)=\sum\limits_{t=j}^{j+L-1}h[X^a(t), X^b(t+\tau)]
\\ \vspace{0.2cm}=\sum\limits_{t=j}^{j+L-1}h[\alpha^{\langle t\rangle_{p^m-1}}+\langle t\rangle_{p}+a,\alpha^{\langle t+\tau \rangle_{p^m-1}}+  \langle t+\tau \rangle_{p} +b]
\\ \vspace{0.2cm}=\sum\limits_{t=j}^{j+L-1}h[a-b-\langle \tau \rangle_p, \alpha^{\langle t\rangle_{p^m-1}}(\alpha^{\langle \tau \rangle_{p^m-1}}-1)].
\end{array}
\]
Denote $\tau_0=\langle \tau \rangle_{p^m-1}$ and $\tau_1=\langle \tau \rangle_p$.
According to the values of $a,b$ and $\tau$, we distinguish four cases.

Case 1: $a=b$ and $\tau_1=0$. In this case $\tau_0\neq 0$ and $\alpha^{\tau_0}-1\in GF(p^m)\setminus \{0\}$. Then \[
\begin{array}{l}
\vspace{0.2cm}H_{X^a,X^a}(\tau:j|L)=\sum\limits_{t=j}^{j+L-1}h[0, \alpha^{\langle t\rangle_{p^n-1}}(\alpha^{\tau_0}-1)]
\\ \vspace{0.2cm}\hspace{2.6cm}=0.
\end{array}
\]

Case 2: $a=b$ and $\tau_1\neq 0$.
If $\tau_0=0$, then
\[
\begin{array}{l}
H_{X^a,X^a}(\tau:j|L)=\sum\limits_{t=j}^{j+L-1}h[-\tau_1, 0]=0.
\end{array}
\]
Otherwise, let $t_0$ be an integer such that $-\tau_1=\alpha^{\langle t_0\rangle_{p^m-1}}(\alpha^{\tau_0}-1)$. Then
\[
\begin{array}{l}
\vspace{0.2cm}H_{X^a,X^a}(\tau :j|L)=\sum\limits_{t=j}^{j+L-1}h[-\tau_1, \alpha^{\langle t\rangle_{p^m-1}}(\alpha^{\tau_0}-1)]
\\ \vspace{0.2cm}\hspace{2.6cm}=\sum\limits_{t=j}^{j+L-1}|\{t:~ t\equiv t_0 \pmod {p^m-1}\}|
\\ \vspace{0.2cm}\hspace{2.6cm}\leq \left\lceil \frac{L}{p^m-1} \right\rceil.
\end{array}
\]

Case 3: $a\neq b$ and $\tau_0=0$. Since $a-b\notin GF(p)$, we have $a-b\neq \tau_1$. Then
$$ H_{X^a,X^b}(\tau :j|L)=\sum\limits_{t=j}^{j+L-1}h[a-b-\tau_1, 0]=0.$$

Case 4: $a\neq b$ and $\tau_0\neq 0$. Clearly, $a-b-\tau_1\neq 0$. Let $t_0$ be an integer such that $a-b-\tau_1=\alpha^{\langle t_0\rangle_{p^m-1}}(\alpha^{\tau_0}-1)$. Then,
\[
\begin{array}{l}
\vspace{0.2cm}
H_{X^a,X^b}(\tau:j|L)
\\ =\sum\limits_{t=j}^{j+L-1}h[a-b-\tau_1,\alpha^{\langle t\rangle_{p^m-1}}(\alpha^{\tau_0}-1)]
\vspace{0.1cm} \\ =\sum\limits_{t=j}^{j+L-1}|\{t:~ t\equiv t_0 \pmod {p^m-1}\}|
\vspace{0.1cm}\\ \leq \left\lceil \frac{L}{p^m-1} \right\rceil.
\end{array}
\]
In summary, the discussion in the four cases above shows that $H(S; L)\leq \left\lceil\frac{L}{p^m-1}\right\rceil$.

Finally, we prove that $H(S;L)\geq \left\lceil \frac{L}{p^m-1} \right\rceil$.
Note that $S$ contains $p^{m-1}$ FHSs of length $p(p^m-1)$ over an alphabet of size $p^m$. Since $I=\lfloor \frac{p(p^m-1)p^{m-1}}{p^m}\rfloor=p^m-1$, by Lemma \ref{optimal set} we get
\[
\begin{array}{l}
H(S; L)\\ \geq \left\lceil \frac{L}{p(p^m-1)} \left\lceil\frac{2(p^m-1)p(p^m-1)p^{m-1}-p^m(p^{m}-1)p^m}{(p(p^m-1)p^{m-1}-1)p^{m-1}}\right\rceil \right\rceil  \vspace{0.2cm}\\
 =\left\lceil \frac{L}{p(p^m-1)} \left\lceil p-\frac{p(2p^m-3)}{(p^{m}-1)p^m-1}\right\rceil \right\rceil \vspace{0.2cm} \\
  =\left\lceil \frac{L}{p^m-1} \right\rceil.
\end{array}
\]
Therefore, $\{X^a:~ a\in R\}$ is a strictly optimal $(p(p^m-1),p^{m-1},p;p^{m})$-FHS set with respect to the bound ({\ref{Bound 6}}).
\end{IEEEproof}

We illustrate the idea of Theorem \ref{optimal p} with $m=2$ and $p=3$ in the following example.
\begin{exmp}
Using the primitive polynomial $f(x)=x^2+x+2\in GF(3)[x]$, we construct the $GF(9)$ as $GF(3)[\alpha]/f(\alpha)$ where $\alpha^2+\alpha+2=0$. The 9 elements of $GF(9)$ can be represented in the form $a_0 + a_1\alpha, a_0, a_1 \in GF(3)$.
The FHS set $S=\{X^0, X^\alpha ,X^{2\alpha}\}$ generated by Construction B is given by
\[
\begin{array}{l}
\vspace{0.2cm}X^0=(1,\alpha+1,2\alpha,2\alpha+2,0,2\alpha+2,\alpha+2,\\
\vspace{0.2cm}\hspace{1cm} \alpha+2,0, \alpha,2\alpha+2,2\alpha+1,2,2\alpha+1,\alpha+1, \\
\vspace{0.2cm}\hspace{1cm} \alpha+1, 2,\alpha+2,2\alpha+1,2\alpha,1,2\alpha,\alpha,\alpha) \\
\end{array}
\]
\[
\begin{array}{l}
\vspace{0.2cm}X^\alpha=(\alpha+1,2\alpha+1,0,2,\alpha,2,2\alpha+2,\\
\vspace{0.2cm}\hspace{1cm} 2\alpha+2,\alpha, 2\alpha,2,1,\alpha+2,1,2\alpha+1, \\
\vspace{0.2cm}\hspace{1cm} 2\alpha+1, \alpha+2,2\alpha+2,1,0,\alpha+1,0,2\alpha,2\alpha) \\
\vspace{0.2cm}X^{2\alpha}=(2\alpha+1,1,\alpha,\alpha+2,2\alpha,\alpha+2,2,\\
\vspace{0.2cm}\hspace{1cm} 2,2\alpha, 0,\alpha+2,\alpha+1,2\alpha+2,\alpha+1,1, \\
\vspace{0.2cm}\hspace{1cm} 1, 2\alpha+2,2,\alpha+1,\alpha,2\alpha+1,\alpha,0,0).
\end{array}
\]
It is readily checked that $S=\{X^0, X^\alpha ,X^{2\alpha}\}$ is a strictly optimal $(24,3,3;9)$-FHS set with respect to the bound ({\ref{Bound 6}}). Such an FHS set is the same as that in Example \ref{24FHS set}.
\end{exmp}

\section{two recursive constructions of strictly optimal FHS sets}

In this section, two recursive constructions are used to construct strictly optimal individual FHSs and FHS sets. The first recursive construction is based on the cyclic difference matrix (CDM).

A $(w, t, 1)$-CDM is a $t \times  w$ matrix $D=(d_{ij})$ ($0 \le i \le t-1$, $0 \le j \le w-1)$ with entries from $\mathbb{Z}_w$ such that, for any two distinct rows $R_{r}$ and $R_h$, the vector difference $R_h-R_{r}$ contains every residue of $\mathbb{Z}_w$ exactly once. It is easy to see that the property
of a difference matrix is preserved even if we add any element of $\mathbb{Z}_w$ to all entries in any row or column of the difference matrix.
Then, without loss of generality, we can assume that all entries in the first row are zero. Such a difference matrix is said to be {\em normalized}. The $(w,t-1,1)$-CDM obtained from a normalized $(w,t,1)$-CDM by deleting the first row is said to be {\em homogeneous}. The existence of a homogeneous $(w,t-1,1)$-CDM is equivalent to that of a $(w,t,1)$-CDM. Observe that difference matrices have been extensively studied. A large number of known $(w, t, 1)$-CDMs are well documented in \cite{CD2007}. In particular, the multiplication table of the prime field $\mathbb{Z}_p$ is a $(p, p, 1)$-CDM. By using the usual product construction of CDMs, we have the following existence result.

\begin{lemma}
\label{CDM}({\rm \cite{CD2007}})
Let $w$ and $t$ be integers with $w\geq t\geq 3$. If $w$ is odd and the least prime factor of $w$ is not less than $t$, then there exists an $(w,t,1)$-CDM.
\end{lemma}


\begin{theorem}\label{difference matrix 2}
Assume that $\{{\cal B}_0,\ldots,{\cal B}_{M-1}\}$ is an $(mg,g,\{K_0,K_1,\ldots,K_{M-1}\},1)$-BNCRDP of size $u$ such that all elements of base blocks of ${\cal B}_j$, together with $0,m,\ldots,(s-1)m,$ form a complete system of representatives for the cosets of $sm\mathbb{Z}_{mg}$ in $\mathbb{Z}_{mg}$  for $0\leq j <M$, where $s|g$ and ${\cal B}_j=\{B_0^j,B_1^j,\ldots,B_{u-1}^j\}$. If there exists a homogeneous $(w, t, 1)$-CDM over $\mathbb{Z}_w$ with $t=\max \limits_{0\leq k <u}\{\sum \limits_{j=1}^M |B_k^j|\}$ and $gcd(w,\frac{g}{s})=1$, then there also exists an $(mgw,gw,\{K_0,K_1,\ldots,K_{M-1}\},1)$-BNCRDP of size $uw$, ${\cal B}'=\{{\cal B}'_0,\ldots,{\cal B}'_{M-1}\}$ such that all elements of base blocks of ${\cal B}_{j}'$, together with $0,m,\ldots,(sw-1)m$, form a complete system of representatives for the cosets of $smw\mathbb{Z}_{mgw}$ in $\mathbb{Z}_{mgw}$ for $0\leq j <M$.

\end{theorem}

\begin{IEEEproof}
Let $\Gamma=(\gamma_{i,j})$ be a homogeneous $(w,t,1)$-CDM over $\mathbb{Z}_w$.
For each collection of the following $M$ blocks:
\[
\begin{array}{l}
\vspace{0.2cm} B_i^0=\{a_{i,0,1},\ldots,a_{i,0,k_0}\},
\\ \vspace{0.2cm} B_i^1=\{a_{i,1,k_0+1},\ldots,a_{i,1,k_1}\},
\\ \vspace{0.2cm} \hspace{1.5cm} \vdots
\\ \vspace{0.2cm} B_i^{M-1}=\{a_{i,M-1,k_{M-2}+1},\ldots,a_{i,M-1,k_{M-1}}\},
\end{array}
\]
where $0\leq i <u$, we construct the following $uw$ new blocks:
\[
\begin{array}{l}
 \vspace{0.2cm}B_{(i,k)}^j=\{a_{i,j,k_{j-1}+1}+mg\gamma_{k_{j-1}+1,k},\ldots,a_{i,j,k_{j}}+mg\gamma_{k_{j},k}\}, \\
 \ \ \ \ \ \ \ \ \ \ \ 0\leq j < M, 0\leq k < w.
\end{array}
\]
Set
\[
\begin{array}{l}
\vspace{0.2cm} {\cal B}'_j=\{B_{(i,k)}^j:~0\leq i<u, 0\leq k<w\}, {\rm and}\\
 {\cal B}'=\{{\cal B}'_j:~0\leq j < M \},
\end{array}
\]
 then the size of ${\cal B}'_j$ is $uw$ for $0\leq j<M$. It is left to show  that ${\cal B}'$ the required $(mgw,gw,\{K_0,K_1,\ldots,K_{M-1}\},1)$-BNCRDP.

Firstly, we show that all elements of base blocks ${\cal B}'_j$, together with $0,m,\ldots,(sw-1)m$, form a complete system of representatives for the cosets of $smw\mathbb{Z}_{mgw}$ in $\mathbb{Z}_{mgw}$. Since $gcd(w,\frac{g}{s})=1$, we have that $\{c\cdot \frac{g}{s}: 0\leq c<w\}\equiv\{0,1,\ldots,w-1\}\pmod w$. It follows that  $\{ms\cdot \frac{cg}{s}: 0\leq c<w\}\equiv \{csm: 0\leq c<w\} \pmod {smw}.$ Clearly, $\bigcup_{0\leq k<w}B_{(i,k)}^j=\bigcup_{z\in B_{i}^j}\{z+cmg:0\leq c<w\}$. Since all elements of base blocks of ${\cal B}_j$, together with $0,m,2m,\cdots,(s-1)m$, form a complete system of representatives for the cosets of $sm\mathbb{Z}_{mg}$ in $\mathbb{Z}_{mg}$, we have that $\bigcup_{0\leq i<u}B_{i}^j\equiv \{0,1,2,\ldots,sm-1\}\setminus \{0,m,\ldots,sm-m\} \pmod {sm}$ and
\[
\begin{array}{l}
\vspace{0.2cm} \bigcup\limits_{0\leq i<u}\bigcup\limits_{0\leq k<w}B_{(i,k)}^j \\
\vspace{0.2cm} =\bigcup\limits_{0\leq i<u}\bigcup\limits_{z\in B_{i}^j}\{z+cmg:0\leq c<w\}\\
\vspace{0.2cm} \equiv \bigcup\limits_{0\leq i<u}\bigcup\limits_{z\in B_{i}^j}\{z+csm:0\leq c<w\} \\
\vspace{0.2cm} \equiv \bigcup\limits_{z\in \mathbf{I}_{sm}\setminus \{0,m,\ldots,sm-m\}}\{z+csm:0\leq c<w\}\\
\vspace{0.2cm} \equiv \mathbf{I}_{snw}\setminus \{0,m,\ldots,smw-m\}\pmod {smw},\\
\end{array}
\]
as desired.

Secondly, we show that each ${\cal B}'_j$ is an $(mgw,gw,K_j,1)$-CRDP.
Since ${\cal B}_j$ is an $(mg,g,K_j,1)$-CRDP, we have $\Delta({\cal B}_j)\subset \mathbb{Z}_{mg}\setminus m\mathbb{Z}_{mg}$. Simple computation shows that
\[
\begin{array}{l}
\Delta({\cal B}'_j)=\bigcup\limits_{0\leq i<u, \atop 0\leq k<w}\Delta(B_{(i,k)}^j) \\
=\bigcup\limits_{0\leq i<u}\{a-b+cmg:~ a\neq b \in B_i^j, \ 0\leq c < w\}\\
=\bigcup\limits_{\tau\in \Delta({\cal B}_j)}(mg\mathbb{Z}_{mgw}+\tau)\subset \mathbb{Z}_{mgw}\setminus m\mathbb{Z}_{mgw}.
\end{array}
\]
It follows that ${\cal B}'_j$ is an $(mgw,gw,K_j,1)$-CRDP.

Finally, we show  $\Delta({\cal B}'_j, {\cal B}'_{j'})\subseteq \mathbb{Z}_{mgw}\setminus m\mathbb{Z}_{mgw}$ for $0\leq j\neq j' <M$.
Since $\Delta({\cal B}_j, {\cal B}_{j'})\subset \mathbb{Z}_{mg}\setminus m\mathbb{Z}_{mg}$, we get
\[
\begin{array}{l}
\Delta({\cal B}'_j, {\cal B}'_{j'})=\bigcup\limits_{0\leq i<u}\bigcup\limits_{0\leq k<w}\Delta(B_{(i,k)}^j, B_{(i,k)}^{j'}) \\
=\bigcup\limits_{0\leq i<u}\{b-a+cmg:~ (a,b) \in B_i^j\times B_i^{j'}, \ 0\leq c < w\}\\
=\bigcup\limits_{\tau\in \Delta({\cal B}_j,{\cal B}_{j'})}(mg\mathbb{Z}_{mgw}+\tau)\subset \mathbb{Z}_{mgw}\setminus m\mathbb{Z}_{mgw}.
\end{array}
\]
Therefore, ${\cal B}'$ is the required BNCRDP.
This completes the proof.
\end{IEEEproof}

The significance of Theorem \ref{difference matrix 2} is that it gives us an effective way to construct a BNCRDP such that all elements of base blocks of each CRDP, together with $0,m,\ldots,(sw-1)m$, form a complete system of representatives for the cosets of $smw\mathbb{Z}_{mgw}$ in $\mathbb{Z}_{mgw}$, which is crucial to our investigation of this paper.

By employing Theorem \ref{difference matrix 2}, we obtain the following construction for resolvable CRDFs.

\begin{corollary}\label{product-resolvable-CRDF}
Assume that there exists a resolvable $(mg,g,k,1)$-CRDF, a resolvable $(gw,g, k, 1)$-CRDF and a $(w, k+1, 1)$-CDM. If gcd$(w, k-1)=1$, then there exists a resolvable $(mgw,g,k,1)$-CRDF.
\end{corollary}
\begin{IEEEproof}
Since there exists a $(w, k+1, 1)$-CDM, there exists a homogeneous $(w, k, 1)$-CDM. Since there exists a resolvable $(mg,g,k,1)$-CRDF and gcd$(w, k-1)=1$, applying Theorem \ref{difference matrix 2} with $M=1$ and $s=\frac{g}{k-1}$ gives a resolvable $(mgw,gw,k,1)$-CRDF ${\cal B}$. Let ${\cal A}$ be a resolvable $(gw, g, k, 1)$-CRDF. Then, ${\cal B}\bigcup \{mA:~ A\in {\cal A}\}$ is a resolvable $(mgw,g,k,1)$-CRDF, where $mA=\{ma:~a\in A\}$.
\end{IEEEproof}

Remark: When $g=k-1$, the recursive construction for CRDFs from Corollary \ref{product-resolvable-CRDF} has been given \cite{JV1984}.

Applying Corollary \ref{product-resolvable-CRDF} with the known resolvable CRDFs in section IV-B, we get the following new resolvable CRDFs.

\begin{corollary}\label{new-CRDF}
(1) There exists a resolvable $(v,4,3,1)$-CRDF for all
$v$ of the form $4p_1p_2\cdots p_u$ where each $p_j$ $\equiv 7 \pmod {12}$ is a prime .

(2) There exists a resolvable $(v,6,3,1)$-CRDF for all
$v$ of the form $6p_1p_2\cdots p_u$ where each $p_j$ $\equiv 5 \pmod {8}$ is a prime.
\end{corollary}

\begin{IEEEproof}
We only prove the first case. The other case can be handled similarly. We prove it by induction on $t$. For $u=1$, by Lemma \ref{4-CRDF} the assertion holds. Assume that the assertion holds for $u=r$ and consider $u=r+1$. Start with a resolvable $(4p_1p_2\cdots p_r,4,3,1)$-CRDF over $\mathbb{Z}_{4p_1p_2\cdots p_r}$  which exists by induction hypothesis. By Lemma \ref{CDM} and Lemma \ref{4-CRDF}, there exists a $(p_{r+1},5,1)$-CDM and a resolvable $(4p_{r+1},4,3,1)$-CRDF. Applying Corollary \ref{product-resolvable-CRDF} yields a resolvable $(4p_1p_2\cdots p_{r+1},4,3,1)$-CRDF. So, the conclusion holds by induction.
\end{IEEEproof}

The application to the CRDFs in Corollary \ref{new-CRDF} gives the following strictly optimal FHSs.
\begin{theorem}\label{ A FHS}

(1) There exists a strictly optimal $(v,2;\frac{v+2}{3})$-FHS for all
$v$ of form $4p_1p_2\ldots p_u$ where each $p_j$ is a prime $\equiv 7 \pmod{12}$.

(2) There exist a strictly optimal $(v,2;\frac{v}{3}+1)$-FHS for all
$v$ of form $6p_1p_2\ldots p_u$ where each $p_j\equiv 5 \pmod{8}$ is a prime.
\end{theorem}

\begin{IEEEproof}
We only prove the second case. The first case can be handled similarly. By Corollary \ref{new-CRDF}, there exists a resolvable $(v,6,3,1)$-CRDF. By Theorem \ref{ strictly optimal 2u} and Theorem \ref{optimal character}, there exists a $(6,2,2)$-CDP, ${\cal A}$ of size $3$ with $d_i^{\cal A}\geq \lfloor \frac{6i}{2}\rfloor=3i$ for $1\leq i\leq 2$. Applying Lemma \ref{CRDP=>CDP} with $g=6$ and $s=3$, there exists a partition-type $(v,\{2,3\},2)$-CDP over $\mathbb{Z}_{v}$ with $d_i\geq \frac{vi}{2}$ for $1\leq i\leq 2$. Applying Theorem \ref{optimal character} with $l=3+\frac{v-6}{3}=\frac{v+3}{3}$, we obtain a strictly optimal $(v, 2; \frac{v+3}{3})$-FHS.
\end{IEEEproof}

By virtue of Lemma \ref{cyclotomic construction}, we can apply Theorem \ref{difference matrix 2} to produce the following series of strictly optimal FHS sets meeting the lower bound ({\ref{Bound 6}}).

\begin{corollary}\label{euv}
Let $v$ be an odd integer of the form $v=p_1^{m_1}p_2^{m_2}\cdots p_s^{m_s}$ for $s$ positive integers $m_1,m_2,\ldots,m_s$ and $s$ primes $p_1,p_2,\ldots,p_s$ with $p_1<p_2<\cdots<p_s$. Let $e$ be a common factor of $p_1-1, p_2-1, \ldots, p_s-1$ such that $2<2e<p_1$ and let $f=\frac{p_1-1}{e}$.
Let $w$ be an odd integer of the form $w=q_1^{n_1}q_2^{n_2}\ldots q_t^{n_t}$ for $t$ positive integers $n_1,n_2,\ldots,n_t$ and $t$ distinct primes $q_1,q_2,\ldots, q_t$ with $q_1<\ldots<q_t$ such that $p_1\leq q_1$ and $gcd(e,w)=1$. Let $r$ be a common factor of $e, q_1-1, q_2-1, \ldots, q_t-1$ with $r>1$. If $v>e(e-2)$, then there exists a strictly optimal $(ewv,f, e; (v-1)w+\frac{ew}{r})$-FHS set with respect to the bound ({\ref{Bound 6}}).
\end{corollary}

\begin{IEEEproof}
Firstly, we prove that there exists a partition-type $(ewv,\{K'_0,\ldots,K'_{f-1}\},e)$-BNCDP of size $(v-1)w+\frac{ew}{r}$, $S'$ such that $d_i^{S'}\geq wvi$ for $1\leq i\leq e$, where $K'_0=\cdots=K'_{f-1}=\{e,r\}$.

By Lemma \ref{cyclotomic construction}, there exists an $(ev,e,\{K_0,\ldots,K_{f-1}\},1)$-BNCRDP of size
$\frac{v-1}{e}$ such that all elements of base blocks of each CRDP, together with $0$, form a complete system of representatives for the cosets of $v\mathbb{Z}_{ev}$ in $\mathbb{Z}_{ev}$ where $K_0=\cdots=K_{f-1}=\{e\}$. Since $p_1\leq q_1$, by Lemma \ref{CDM} there exists a homogeneous $(w, p_1-1, 1)$-CDM over $\mathbb{Z}_w$. Since $gcd(e,w)=1$, applying Theorem \ref{difference matrix 2} with $g=e,s=1$ yields an $(ewv,ew,\{K_0,K_1,\ldots,K_{f-1}\},1)$-BNCRDP of size $\frac{(v-1)w}{e}$ such that all elements of base blocks of each CRDP, together with $0,v,\ldots,(w-1)v$, form a complete system of representatives for the cosets of $wv\mathbb{Z}_{ewv}$ in $\mathbb{Z}_{ewv}$. By Corollary \ref{vu}, there exists a partition-type $(ew,\{K''_0,\ldots,K''_{f-1}\},e)$-BNCDP of size $\frac{ew}{r}$, $S$ with $d_i^S\geq wi$ for $1\leq i\leq e$ where $K''_0=\cdots=K''_{f-1}=\{r\}$. By applying Lemma \ref{CRDPs=>CDPs} with $g=ew$ and $s=w$ we obtain a partition-type $(evw,\{K'_0,\ldots,K'_{f-1}\},e)$-BNCDP of size $(v-1)w+\frac{ew}{r}$, $S'$ with $d_i^{S'}\geq wvi$ for $1\leq i\leq e$, where $K'_0=\cdots=K'_{f-1}=\{e,r\}$.

Finally, we show $\left\lceil\frac{2InM-(I+1)Il}{(nM-1)M}\right\rceil=e$.

By definition, we have
\[
\begin{array}{l}
\vspace{0.2cm}I=\left\lfloor \frac{evw \cdot \frac{p_1-1}{e}}{(v-1)w+\frac{ew}{r}} \right \rfloor \\
\vspace{0.2cm}\hspace{0.3cm} =\left\lfloor \frac{rv(p_1-1)}{(v-1)r+e} \right \rfloor \\
\vspace{0.2cm}\hspace{0.3cm} =
\left\{\begin{array}{ll}
p_1-1 & {\rm  if} \ \ r=e, \ {\rm and} \\
p_1-2  & {\rm  if} \ \ r< e.\
\end{array}
\right .
\end{array}
\]
If $r=e$, we have
\[
\begin{array}{l}
\vspace{0.2cm}\left\lceil\frac{2InM-(I+1)Il}{(nM-1)M}\right\rceil =\left\lceil\frac{2(p_1-1)ewvf-p_1(p_1-1)wv}{(ewvf-1)f}\right\rceil
\\\vspace{0.2cm} \hspace{2.5cm}= \left\lceil\frac{e(p_1-2)wv}{wv(p_1-1)-1}\right\rceil \\
\vspace{0.2cm}\hspace{2.5cm}= \left\lceil e-\frac{e(wv-1)}{wv(p_1-1)-1}\right\rceil \\
\hspace{2.5cm} =e.
\end{array}
\]
Otherwise, we have
{\small \begin{equation}
\label{Bound e}
\begin{aligned}
&\left\lceil\frac{2InM-(I+1)Il}{(nM-1)M}\right\rceil \\ &=  \left\lceil\frac{2(p_1-2)ewv\frac{p_1-1}{e}-(p_1-1)(p_1-2)(wv-w+\frac{ew}{r})}{(ewv\frac{p_1-1}{e}-1)\frac{p_1-1}{e}}\right\rceil
\\ 
& = \left\lceil e-\frac{wve+\frac{(p_1-2)we(e-r)}{r}-e}{wv(p_1-1)-1}\right\rceil.
\end{aligned}
\end{equation}
}
Since $p_1>2e$, $e\geq r\geq 2$ and $v>e(e-2)$, we have $v(p_1-1)>ve+\frac{(p_1-2)e(e-r)}{r}$, which leads to $$0<\frac{wve+\frac{(p_1-2)we(e-r)}{r}-e}{wv(p_1-1)-1}<1.$$
It follows from (\ref{Bound e}) that $\left\lceil\frac{2InM-(I+1)Il}{(nM-1)M}\right\rceil=e$. By Theorem \ref{optimal sets character}, $S'$ is a strictly optimal $(ewv,f, e; (v-1)w+\frac{ew}{r})$-FHS set with respect to the bound ({\ref{Bound 6}}). This completes the proof.
\end{IEEEproof}
Now, we present the second recursive construction.

{\bf Construction C}:
Let $v$ be an odd integer of the form $v=p_1^{m_1}p_2^{m_2}\cdots p_s^{m_s}$ for $s$ positive integers $m_1,m_2,\ldots,m_s$ and $s$ distinct primes $p_1,p_2,\ldots,p_s$. Let $e$ be a common factor of $p_1-1, p_2-1, \ldots, p_s-1$ with $e>1$, and let $f=\min\{\frac{p_i-1}{e}~:1\leq i\leq s\}$. Let $\{X_0,X_1,\ldots, X_{M-1}\}$ be an $(e,M,\lambda;l)$-FHS set over $F$, where $X_i=\{x_i(t)\}_{t=0}^{e-1}$ and $M\leq f$. For $0\leq i< M$, let $Y_i = \{y_i(t)\}_{t=0}^{ve-1}$ be the FHS over $\mathbb{Z}_v\times F$, defined by
$$y_i(t)= (\langle a^ig^tt\rangle_v, x_i(\langle t\rangle_e))$$
where $a$ and $g$ are defined in Section IV-C,
 $\langle z\rangle_u$ denotes the least nonnegative residue of $z$ modulo $u$ for any positive integer $u$ and any integer $z$.

\begin{lemma}\label{correlation connection}
Let $Y_i$ and $Y_j$ be any two FHSs in Construction C. Then for $ 0\leq \tau, c <ve$ and $1\leq L \leq ve$, \ it holds that $H_{Y_i,Y_j}(\tau; c|L)\leq H_{X_i,X_j}(\langle\tau \rangle_e; k|\lceil \frac{L}{v} \rceil)$ for some $k$ with $0\leq k <e$.
\end{lemma}
\begin{IEEEproof}
By definition,
{  \[
\begin{array}{l}
\vspace{0.2cm}h[y_i(t),y_j(t+\tau)]\\
\vspace{0.2cm}=h[(\langle a^ig^tt\rangle_v, x_i(\langle t\rangle_e)), (\langle a^jg^{(t+\tau)}(t+\tau)\rangle_v, x_j(\langle t+\tau\rangle_e))] \\
\vspace{0.2cm}=h[\langle a^ig^tt\rangle_v, \langle a^jg^{(t+\tau)}(t+\tau)\rangle_v]\cdot h[x_i(\langle t\rangle_e), x_j(\langle t+\tau\rangle_e)]\\
=h[\langle(a^i-a^jg^{\tau})t\rangle_v, \langle a^jg^{\tau}\tau \rangle_v]\cdot h[x_i(\langle t\rangle_e), x_j(\langle t+\tau\rangle_e)].
\end{array}
\]
}
According to the parameters $i,j,\tau$, we distinguish two cases.

Case 1: $i=j$ and $\langle\tau \rangle_e=0$. Since $g^{\tau}=1$ and $x_i(\langle t\rangle_e)=x_j(\langle t+\tau\rangle_e)$, we get
{\small  \[
\begin{array}{l}
\vspace{0.3cm}H_{Y_i,Y_j}(\tau;c|L)=\sum\limits_{t=c}^{c+L-1}h[y_i(t), y_j(t+\tau)] \\
\vspace{0.3cm}=\sum\limits_{t=c}^{c+L-1}h[0, \langle a^jg^{\tau}\tau \rangle_v]\\
\vspace{0.3cm}=
\begin{array}{l}
\left\{\begin{array}{ll}
L  & {\rm if} \ \tau=0 ,\\
0  & {\rm otherwise}.
\end{array}
\right .
\end{array}
\end{array}
\]}
The last equality holds since $gcd(v,e)=1$. Therefore, $H_{Y_i,Y_j}(\tau; c|L) \leq H_{X_i,X_j}(\langle\tau \rangle_e; k|\lceil \frac{L}{v} \rceil)$ for $0\leq k <e$.

Case 2: $i\neq j$ or $\langle\tau \rangle_e\neq 0$. In this case, $a^i-a^jg^{\tau} \in U(\mathbb{Z}_v)$.  Set
\[
\begin{array}{l}
\vspace{0.2cm}t_0\equiv a^jg^{\tau}\tau\cdot (a^i-a^jg^{\tau})^{-1} \pmod v.
\end{array}
\]
Then $h[\langle(a^i-a^jg^{\tau})t\rangle_v, \langle a^jg^{\tau}\tau \rangle_v]=1$ if and only $\langle t\rangle_v=\langle t_0\rangle_v$.
Since $e|(p_r-1)$ for $0<r\leq s$, it holds that $v\equiv 1\pmod e$. Then
{ \footnotesize \[
\begin{array}{l}
\vspace{0.3cm}H_{Y_i,Y_j}(\tau;c|L)=\sum\limits_{t=c}^{c+L-1}h[y_i(t), y_j(t+\tau)] \\
\vspace{0.3cm}=\sum\limits_{t=c}^{c+L-1}h[\langle(a^i-a^jg^{\tau})t\rangle_v, \langle a^jg^{\tau}\tau \rangle_v]\cdot h[x_i(\langle t\rangle_e), x_j(\langle t+\tau\rangle_e)]\\
\vspace{0.3cm}=\sum\limits_{ 0\leq a<e \atop t_0+av \in \{c,c+1,\ldots,c+L-1\} } h[x_i(\langle t_0+av\rangle_e), x_j(\langle t_0+av+\tau\rangle_e)]\\
\vspace{0.3cm}=\sum\limits_{a= \lceil \frac{c-t_0}{v} \rceil}^{\lfloor \frac{c+L-1-t_0}{v} \rfloor} h[x_i(\langle t_0+a\rangle_e), x_j(\langle t_0+a+\tau\rangle_e)]\\
\vspace{0.3cm}\leq H_{X_i,X_j}(\langle\tau \rangle_e; t_0+\lceil \frac{c-t_0}{v} \rceil|\lceil \frac{L}{v} \rceil).
\end{array}{}
\]
}
This completes the proof.
\end{IEEEproof}

\begin{theorem}\label{optimal(2)}
Let $v$ be an odd integer of the form $v=p_1^{m_1}p_2^{m_2}\cdots p_s^{m_s}$ for $s$ positive integers $m_1,m_2,\ldots,m_s$ and $s$ distinct primes $p_1,p_2,\ldots,p_s$. Let $e$ be a positive integer such that $e|p_i-1$ for $1\leq i\leq s$. If there is a strictly optimal $(e,\lambda;l)$-FHS with respect to the bound ({\ref{Bound 4}}) over $F$ such that $\lambda|e$ and $e>l$, then the sequence $Y$ in Construction C is a strictly optimal $(ve,\lambda;vl)$-FHS with respect to the bound ({\ref{Bound 4}}).
\end{theorem}
\begin{IEEEproof}
We first prove that $Y$ in Construction C is a $(ve,\lambda; vl)$-FHS. Let $X$ be a strictly optimal $(e,\lambda;l)$-FHS, we have $\lambda l\leq e< (\lambda+1)l$. Let $e=\lambda l+\epsilon,\ 0\leq \epsilon<l$, then $ve=\lambda(vl)+v\epsilon$.
Since $X$ meets the bound ({\ref{Bound 4}}), by Lemma \ref{FHS BOUND} we get
\begin{equation}
\label{connection a X}
H(X;L')=\left\lceil\frac{L'}{e}\left\lceil \frac{(e-\epsilon)(e+\epsilon-l)}{l(e-1)} \right\rceil \right\rceil = \left\lceil \frac{\lambda L'}{e}\right\rceil
\end{equation}
for $1\leq L' \leq e$. By Lemma \ref{correlation connection}, we have $H(Y,ve)=\max\limits_{1\leq \tau< ve}\{H_{Y,Y}(\tau; 0|ve)\}\leq \max\limits_{1\leq \tau< ve}\{H_{X,X}(\langle\tau\rangle_e; 0|e)\}= \lambda$.  By Lemma \ref{A optimal}, we have $H(Y,ve)= \lambda$, i.e,  $Y$ is a $(ve,\lambda; vl)$-FHS.

It remains to prove that $Y$ is strictly optimal with respect to the bound ({\ref{Bound 4}}). On one hand, by Lemma {\ref{FHS BOUND}}, we have that
\[
\begin{array}{l}
\vspace{0.3cm}H(Y;L)\geq \left\lceil\frac{L}{ve}\left\lceil \frac{(v(\lambda l+\epsilon)-v\epsilon)(v(\lambda l+\epsilon)+v\epsilon-vl)}{vl(v(\lambda l+\epsilon)-1)} \right\rceil \right\rceil \\
\vspace{0.3cm}\hspace{1.3cm}=\left\lceil\frac{L}{ve}\left\lceil\lambda- \frac{\lambda vl-\lambda v\epsilon-\lambda}{v\lambda l+v\epsilon-1} \right\rceil \right\rceil \\
\vspace{0.3cm}\hspace{1.3cm}=\left\lceil\frac{\lambda L}{ve} \right\rceil
\end{array}{}
\]
for $1\leq L\leq ve.$
On the other hand, by Lemma {\ref{correlation connection}}, equality ({\ref{connection a X}}) and the fact that $\frac{e}{\lambda}$ is an integer, we have

\[
\begin{array}{l}
\vspace{0.3cm}H(Y;L)\leq H(X;\lceil \frac{L}{v} \rceil)=\left\lceil\frac{\lambda}{e} \lceil \frac{L}{v}\rceil\right\rceil=\left\lceil\frac{1}{\frac{e}{\lambda}} \lceil \frac{L}{v}\rceil\right\rceil= \left\lceil\frac{L\lambda}{ve} \right\rceil,
\end{array}{}
\]
where the last equality holds because of the property of ceiling function given in page 71 of {\cite{GKP1994}}. Therefore, $H(Y;L)=\left\lceil\frac{\lambda L}{ve} \right\rceil$.
This completes the proof.
\end{IEEEproof}

\begin{theorem}\label{optimal(4)}
Let $v$ be an odd integer of the form $p_1^{m_1}p_2^{m_2}\cdots p_s^{m_s}$ for $s$ positive integers $m_1,m_2,\ldots,m_s$ and $s$ primes $p_1,p_2,\ldots,p_s$ with $p_1<p_2<\cdots<p_s$. Let $e>1$ be an integer such that $e|p_i-1$ for $0<i\leq s$, and $f=\min\{\frac{p_i-1}{e}~:1\leq i\leq s\}$. Let $M$ be positive integer such that $M\leq f$. If there is  a strictly optimal $(e,M,\lambda;l)$-FHS set with respect to the bound ({\ref{Bound 6}}) over $F$, ${\cal A}=\{X_0,X_1,\ldots, X_{M-1}\}$, such that $\lambda|e$ and $e>l\geq 2$, then ${\cal B}$=$\{Y_0,Y_1,\ldots, Y_{M-1}\}$ generated by Construction C is a strictly optimal $(ve,M,\lambda;vl)$-FHS set with respect to the bound ({\ref{Bound 6}}).
\end{theorem}
\begin{IEEEproof}
Since ${\cal A}$ meets the bound ({\ref{Bound 6}}),  ${\cal A}$ is an optimal FHS set and $H({\cal A};e)=\lambda$. By Lemma {\ref{correlation connection}}, we have $H({\cal B};ve)\leq H({\cal A};e)=\lambda$. By Lemma \ref{set}, we have that ${\cal B}$ is a $(ve,M,\lambda; vl)$-FHS set.

It remains to prove that ${\cal B}$ is a strictly optimal FHS set with respect to the bound ({\ref{Bound 6}}). By the definition of $I$ in Lemma {\ref{optimal set}}, we have $I=\left\lfloor \frac{eM}{l}\right\rfloor$. Set $eM=Il+r$ where $r$ is the least nonnegative residue module $l$. Since ${\cal A}$ is a strictly optimal $(e,M,\lambda;l)$-FHS set with respect to the bound ({\ref{Bound 6}}), we get
\begin{equation}
\label{connection Xs}
H({\cal A};L)=\left\lceil \frac{L}{e} \left\lceil\frac{2IeM-(I+1)Il}{(eM-1)M}\right\rceil \right\rceil=\left\lceil\lambda\cdot \frac{L}{e} \right\rceil,
\end{equation}
for $1\leq L\leq e$. Obviously, $-\frac{1}{M}< \frac{I(r+1-l)}{(Il+r-1)M} \leq 0$. Therefore,
{\small
\begin{equation}
\label{connection 1}
\lambda = \left\lceil\frac{I(Il+2r-l)}{(Il+r-1)M}\right\rceil=\left\lceil\frac{I}{M}+\frac{I(r+1-l)}{(Il+r-1)M}\right\rceil=\left\lceil\frac{I}{M}\right\rceil
\end{equation}
}
On one hand, by Lemma {\ref{correlation connection}} and equality ({\ref{connection Xs}}), we have
\begin{equation}
\label{connection X,Y}
\vspace{0.3cm}H({\cal B};L)\leq H({\cal A}; \left\lceil \frac{L}{v} \right\rceil)=\left\lceil\frac{\lambda}{e}\cdot \left\lceil \frac{L}{v}\right\rceil\right\rceil=\left\lceil\frac{ L}{ve}\cdot \lambda\right\rceil,
\end{equation}
where the last equality holds by the property of ceiling function given in page 71 of  {\cite{GKP1994}} since $\frac{e}{\lambda}$ is an integer.
On the other hand, by Lemma \ref{optimal set}, we get

{ \begin{equation}
\label{Y Bound}
\begin{aligned}
 H({\cal B};L) &\geq \left\lceil \frac{L}{ve} \left\lceil\frac{2I(ve)M-(I+1)I(vl)}{((ve)M-1)M}\right\rceil \right\rceil \\
 &=\left\lceil \frac{L}{ve} \left\lceil\frac{Iv(Il+2r-l)}{(Ivl+vr-1)M}\right\rceil \right\rceil \\
 &=\left\lceil \frac{L}{ve} \left\lceil \frac{I}{M}+\frac{ Irv-vIl+I}{(vIl+rv-1)M} \right\rceil \right\rceil\\
 &=\left\lceil \frac{L}{ve} \cdot \left\lceil \frac{I}{M} \right\rceil \right\rceil\\
 &=\left\lceil \frac{L}{ve} \cdot \lambda \right\rceil,
\end{aligned}
\end{equation}
}
where the last equality holds because of equality (\ref{connection 1}).

From equalities (\ref{connection X,Y}) and (\ref{Y Bound}), we have $ H({\cal B};L)= \left\lceil \frac{L}{ve}\cdot\lambda \right\rceil $.
Therefore, ${\cal B}$ is a strictly optimal FHS set with respect to the bound ({\ref{Bound 6}}). This completes the proof.
\end{IEEEproof}

\begin{theorem}\label{known result}(\cite{ZTNP2012})
Let $d, m$ be positive integers with $m\geq 2$ and let $q$ be a prime prower such that $d|q-1$. Then there is a strictly optimal $(\frac{q^m-1}{d},d,\frac{q-1}{d};q^{m-1})$-FHS set $S$, and $H(S;L)=\left\lceil \frac{L(q-1)}{q^m-1}\right\rceil$ for $1\leq L \leq \frac{q^m-1}{d}$.
\end{theorem}

By applying Theorem \ref{optimal(4)} with $e=\frac{q^m-1}{d}, M=d$ and $\lambda=\frac{q-1}{d}$ to Theorem \ref{known result}, we can obtain the following corollary.
\begin{corollary}\label{d,m}
Let $d,m$ be positive integers with $m>1$ and let $q$ be a prime power such that $d|q^m-1$. Let $m_1,m_2,\ldots,m_s$ be $s$ positive integers. Let $p_1,p_2,\ldots,p_s$ be $s$ distinct primes such that $\frac{q^m-1}{d}|p_i-1$ and $q^m\leq p_i$ for all $1\leq i\leq s$. Set $v=p_1^{m_1}p_2^{m_2}\cdots p_s^{m_s}$. Then there exists a strictly optimal $(v\frac{q^m-1}{d},d,\frac{q-1}{d};vq^{m-1})$-FHS set with respect to the bound ({\ref{Bound 6}}).
\end{corollary}
\section{Concluding Remarks}  %
\label{concl}                               %

In this paper, a combinatorial characterization of strictly optimal FHSs and FHS set was obtained.  Some new individual FHSs and FHS sets having strictly optimal Hamming correlation with respect to the  bounds were presented.  It would be nice if more individual FHSs and FHS sets whose partial Hamming correlation achieves the lower bounds could be constructed. It may be possible that some lower bounds on the partial Hamming correlation of FHSs could be improved from the combinatorial characterization.

\ifCLASSOPTIONcaptionsoff
  \newpage
\fi

\end{document}